\pdfoutput=1
\documentclass[aps, reprint, superscriptaddress, letterpaper, floatfix, prb]{revtex4-1}
\usepackage{graphicx}
\usepackage{color}
\usepackage{subfigure}
\usepackage{tabularx}
\usepackage{floatrow}
\usepackage{dcolumn}
\usepackage{bm}
\usepackage{amsmath}
\usepackage{scrextend}
\usepackage{hyperref}

\begin{document}

\title{Band-to-band transitions, selection rules, effective mass and exciton binding energy parameters in monoclinic $\beta$-Ga$_2$O$_3$} 

\author{Alyssa Mock}
\email[Electronic mail: ]{amock@huskers.unl.edu}
\affiliation{Department of Electrical and Computer Engineering and Center for Nanohybrid Functional Materials, University of Nebraska-Lincoln, Lincoln, NE 68588, USA}
\author{Rafa\l{} Korlacki}
\affiliation{Department of Electrical and Computer Engineering and Center for Nanohybrid Functional Materials, University of Nebraska-Lincoln, Lincoln, NE 68588, USA}
\author{Chad~Briley}
\affiliation{Department of Electrical and Computer Engineering and Center for Nanohybrid Functional Materials, University of Nebraska-Lincoln, Lincoln, NE 68588, USA}
\author{Vanya~Darakchieva}
\affiliation{Department of Physics, Chemistry, and Biology (IFM), Link{\"o}ping University, SE 58183, Link{\"o}ping, Sweden}
\author{Bo~Monemar}
\affiliation{Department of Physics, Chemistry, and Biology (IFM), Link{\"o}ping University, SE 58183, Link{\"o}ping, Sweden}
\affiliation{Global Innovation Research Organization, Tokyo University of Agriculture and Technology, Koganei, Tokyo, Japan}
\author{Y.~Kumagai}
\affiliation{Department of Applied Chemistry, Tokyo University of Agriculture and Technology, Koganei, Tokyo, Japan}
\author{K.~Goto}
\affiliation{Department of Applied Chemistry, Tokyo University of Agriculture and Technology, Koganei, Tokyo, Japan}
\affiliation{Tamura Corporation, Sayama, Saitama, Japan}
\author{M.~Higashiwaki}
\affiliation{National Institute of Information and Communications Technology, Koganei, Tokyo, Japan}
\author{Mathias Schubert}
\affiliation{Department of Electrical and Computer Engineering and Center for Nanohybrid Functional Materials, University of Nebraska-Lincoln, Lincoln, NE 68588, USA}
\affiliation{Department of Physics, Chemistry, and Biology (IFM), Link{\"o}ping University, SE 58183, Link{\"o}ping, Sweden}
\affiliation{Leibniz Institute for Polymer Research, Dresden, Germany}

\date{\today}

\begin{abstract}

We employ an eigen polarization model including the description of direction dependent excitonic effects for rendering critical point structures within the dielectric function tensor of monoclinic $\beta$-Ga$_2$O$_3$ yielding a comprehensive analysis of generalized ellipsometry data obtained from 0.75~eV--9~eV. The eigen polarization model permits complete description of the dielectric response, and we obtain single-electron and excitonic band-to-band transition anisotropic critical point structure model parameters including their polarization eigenvectors within the monoclinic lattice. We compare our experimental analysis with results from density functional theory calculations performed using a recently proposed Gaussian-attenuation-Perdue-Burke-Ernzerhof hybrid density functional, and we present and discuss the order of the fundamental direct band-to-band transitions and their polarization selection rules, the electron and hole effective mass parameters for the three lowest band-to-band transitions, and their exciton binding energy parameters, in excellent agreement with our experimental results. We find that the effective masses for holes are highly anisotropic and correlate with the selection rules for the fundamental band-to-band transitions, where the observed transitions are polarized closely in the direction of the lowest hole effective mass for the valence band participating in the transition.

\end{abstract}

\maketitle

\section{Introduction}

\begin{figure}[hbt]
\includegraphics[width=.99\linewidth]{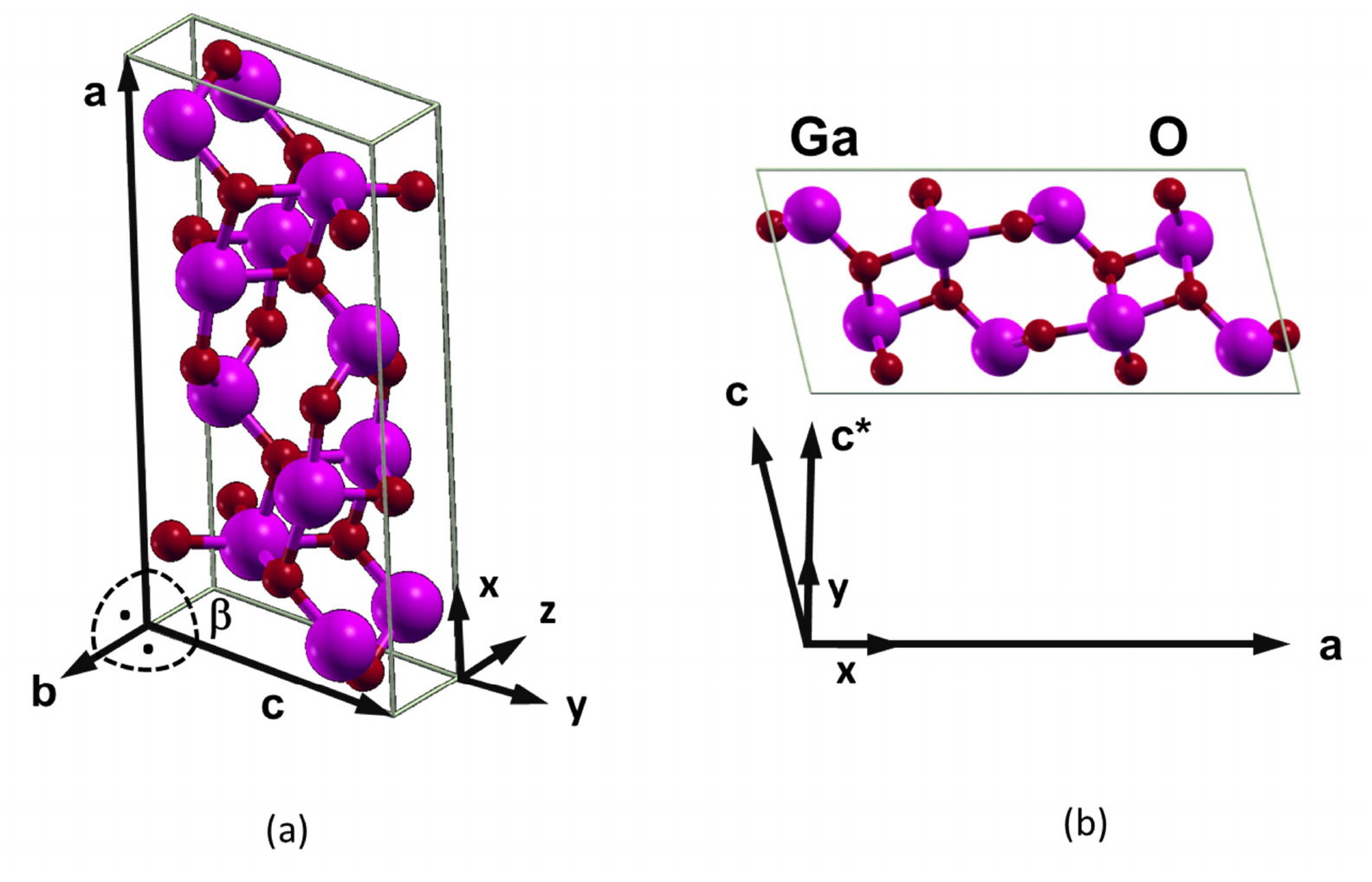}
\caption{(a) Unit cell of $\beta$-Ga$_2$O$_3$ detailing crystallographic and laboratory coordinate systems. (b) Description of the orthogonal laboratory coordinate system within the monoclinic \textbf{a-c} plane. Axis \textbf{c*} is chosen for convenience parallel to laboratory axis \textit{y}, and orthogonal to both \textbf{a} and \textbf{b}.}
\label{unitcell}
\end{figure}

Single crystalline group-III sesquioxides are currently at the forefront of research for applications in electronic and optoelectronic devices due to unique physical properties. Such conductive oxides, including tin doped In$_2$O$_3$ or Ga$_2$O$_3$, can be utilized as transparent thin film electrodes for various devices such as photovoltaic cells\cite{Granqvist_1995}, flat panel displays\cite{Betz_2006}, smart windows\cite{Granqvist_1995,Gogova_1999}, and sensors\cite{Reti_1994}. The highly anisotropic monoclinic $\beta$-gallia crystal structure ($\beta$ phase) is the most stable crystal structure among the five phases ($\alpha$, $\beta$, $\gamma$, $\epsilon$, and $\delta$) of Ga$_2$O$_3$ (Fig. \ref{unitcell})\cite{Roy_1952,Tippins_1965}. It belongs to the space group 12 and has base center monoclinic lattice. Ga$_2$O$_3$ shows potential for use in transparent electronics and high energy photonic applications due to its large band gap of 4.7-4.9 eV\cite{Wager_2003,Sturm_2016,Sturm_2015,Furthmuller_2016}.

Precise and accurate knowledge of the band gap energies, their polarization selection rules and energetic order, and the resulting anisotropy in the dielectric function is a prerequisite to correct assignment of basic physical properties in low symmetry materials. Electronic band-to-band transitions cause critical point (CP) features in the joint density of states, which result in CP structures in the dielectric function\cite{Yubook2005}. Accurate parametric models yield access to CP parameters such as transition energies and polarization selection rules, which allow for direct comparison with results both from experiment, e.g., optical absorption and reflectance measurements, as well as from theory, e.g., density functional theory (DFT) band structure calculations. Additionally, the accurate and explicit knowledge of the spectral dependence of the dielectric function tensor of monoclinic $\beta$-Ga$_2$O$_3$ is equally important as it will allow for accurate and precise prediction of optical and optoelectronic device performances. A suitable and precise technique to determine the complex dielectric function of a material without a-priori implication of a physical model is ellipsometry\cite{Fujiwara_2007}. An extension of this approach permitting for the appropriate concept to acquire experimental data from arbitrarily anisotropic materials has been developed recently by generalized spectroscopic ellipsometry (GSE)\cite{Schubert_1996,Schubert_2006,Fujiwara_2007}. This concept has been used for accurate determination of the spectral dependencies of dielectric function tensor elements for a wide variety of materials with all known crystal symmetries\cite{Schubert_1996,Schubert_2000,Ashkenov_2003,Schubert_2004,Schubert_2004_2,Dressel_2008,ArteagaAO2009,Hofmann_2013,QiAPLpentacene2013,Schubert_2016,MockPRB2017CdWO4}.

The CP model approach was used frequently in the past to model dielectric function spectra as model dielectric function (MDF) approach. Within the MDF, individual contributions corresponding to, for example, band-to-band transitions and associated excitonic contributions, and contributions from spectral regions outside a given spectrum under consideration, sum to the overall dielectric response. The CP-MDF approach was also used successfully to explain anisotropy for many different types of materials, e.g., with tetragonal, hexagonal, and orthorhombic crystal symmetries in single crystal bulk and thin film heterostructures\cite{Schubert_1996,Schubert_2000,Schubert_2004_2,Dressel_2008,Schubert_2004,Ashkenov_2003,Kasic_2000,Hofmann_2013}. Different model functions exist for different types of electronic excitations, e.g., single electronic transitions characterizable by the transition's Van-Hoofe singularity, and displacement and losses caused by many-body interactions (excitons)\cite{Klingshirnbook,Yubook2005,GrundmannPoSbook}. Functions for electronic transitions in the parabolic band approximation, for example, are described by Adachi, Kimura, and Suzuki\citep{AdachiJAPCdTe1993}. Due to the complexity of the potential distribution in many body problems, excitonic contributions are often considered by Lorentzian or Gaussian broadened oscillators. Tangyu provided explicit solutions for Wannier-type excitonic contributions including all bound and unbound states in the parabolic band approximation\cite{Tanguy_1995,TangyuPRL1996,Tanguy_1999}.

Dielectric anisotropic materials can be described by a symmetric dielectric tensor consisting of nine generally independent complex valued elements -- complex valued wavelength dependent functions. In general, these dielectric function tensor elements will display variations with wavelength reflecting the underlying physical mechanisms responsible for the dielectric response of a given material. For example, at long wavelengths the dielectric response is governed by long wavelength active phonon modes and free charge carrier excitations. For near-infrared (NIR), visible (VIS), ultra violet (UV) to vacuum ultra violet (VUV) photon energies, the dielectric response is caused by electronic band-to-band transitions and associated excitonic absorption processes.  For cubic, tetragonal, trigonal, hexagonal, and orthorhombic crystal symmetries, physically meaningful transformations (spatial rotations with real valued angular arguments) can be found independent on wavelength (photon energy) which diagonalize the dielectric function tensor. Jellison \textit{et al.}, investigating the tensor elements of single crystals of monoclinic CdWO$_4$ by GSE in the NIR-UV spectral region, pointed out the necessity to include four independent dielectric tensor elements in order to describe a material with monoclinic crystal symmetry within spectral regions of absorption, equivalent to the fact that no physically meaningful rotations can be found for diagonalization independent of wavelength\cite{Jellison_2011}. This fact has long been known for long wavelength analysis of materials with monoclinic and triclinic crystal symmetries. Born and Huang developed a microscopic description of the long wavelength lattice vibrations in the harmonic approximation, where the inter-atomic forces are considered constant and the equations of motion are determined by harmonic potentials\cite{Born_1954}. In the Born and Huang model, the eigen dielectric displacement modes (phonons) are characterized by a transition frequency (energy), amplitude and displacement eigen vector direction. The Born and Huang model was suggested for long wavelength reflectance dispersion analysis\cite{EmslieJOSA1983}, and used successfully for transverse phonon mode analysis in materials with monoclinic (LiAlSi$_2$O$_6$\cite{PavinichOS1978},
MgCaSi$_2$O$_6$\cite{BelousovOS1978}, CuO\cite{KuzmenkoPRBCuO2001}, MnWO$_4$\cite{MoellerPRBMnWO42014}, Y$_2$SiO$_5$\cite{HoeferVSYSiO52016}, $\beta$-Ga$_2$O$_3$\cite{Schubert_2016}, CdWO$_4$\cite{MockPRB2017CdWO4}) and triclinic crystal symmetries (K$_2$Cr$_2$O$_7$\cite{HoferVS2014}, CuSO$_4 \cdot$ 5H$_2$O\cite{HoferVS2013}). Recently, the extension of the Born and Huang model for the determination of longitudinal phonon modes including free charge carrier properties was derived and demonstrated successfully for $\beta$-Ga$_2$O$_3$\cite{Schubert_2016} and CdWO$_4$\cite{MockPRB2017CdWO4}. A generalization of the Lyddane-Sachs-Teller relation\cite{LyddanePRB1941} in the Born and Huang model to materials with monoclinic and triclinic symmetries was found  excellently satisfied\cite{SchubertPRL2016}.

Fundamental band-to-band transitions in $\beta$-Ga$_2$O$_3$ have been investigated very recently using a variety of techniques including density functional theory (DFT) calculations\cite{Sturm_2015,Furthmuller_2016,Ricci_2016,varley2010,navarro2015}, optical absorption\cite{Ricci_2016}, reflection\cite{Matsumoto_1974,Onuma_2015}, and ellipsometry\cite{Sturm_2015,Sturm_2016}. Due to monoclinic symmetry, all band-to-band transitions are expected to depend on polarization. The eigen polarization of the transitions may not necessarily align with any of the high symmetry crystal axes. In addition, the formation of excitons, upon the optical excitation of a band-to-band transition, strongly modifies the frequency dependence of the dielectric response in semiconductors\cite{Yubook2005}. Thus, in order to accurately determine the transition energies, the excitonic contribution must be accounted for. For $\beta$-Ga$_2$O$_3$, there has been significant discrepancies in reported properties of the fundamental band-to-band transitions. Ricci~\textit{et al.}, ignoring excitonic effects, recently showed optical absorption anisotropy in $\beta$-Ga$_2$O$_3$ with the lowest onset of absorption occurring with polarization in the \textbf{a-c} plane at 4.5-4.6~eV\cite{Ricci_2016}. For polarization along the crystal axis \textbf{b} the absorption onset was unambiguously shifted by 0.2~eV towards shorter wavelength. Onuma \textit{et al.} investigated polarized transmittance and reflectance spectra\cite{Onuma_2015}. As a result of their investigations, an indirect gap band-to-band transition around 4.43~eV and a direct gap transition around 4.48~eV parallel to the \textbf{c} axis were proposed, without considering excitonic effects. Sturm \textit{et al.} considered contributions from both bound and unbound  Wannier-type excitons\cite{Tanguy_1995,Tanguy_1996,Tanguy_1999} and reported the lowest direct gap band-to-band transition at approximately 4.88~eV, polarized within the \textbf{a-c} plane nearly parallel to \textbf{c}. Furthm\"uller and Bechstedt presented quasiparticle band structures and density of states of $\beta$-Ga$_2$O$_3$ obtained from DFT combined with Hedin's \textit{GW} approximation for single-particle excitations\cite{Furthmuller_2016}. The lowest transition energy was determined by this approach to be around 5.04~eV, with polarization mainly along the \textbf{c} axis in the monoclinic \textbf{a-c} plane.

Setyawan and Curtarolo\cite{setyawan2010} demonstrated that materials with base center monoclinic lattices exhibit five distinct first Brillouin zones. Recently, Peelaers and Van de Walle\cite{Peelaerspssb2015Ga2O3meff} discussed the proper Brillouin zone and band structure of $\beta$-Ga$_2$O$_3$ following the recommendations of Setyawan and Curtarolo.\cite{setyawan2010} High-symmetry points and lines connecting points in the Brillouin zone are of general interest for identifying physical properties of materials. Band structure computations for $\beta$-Ga$_2$O$_3$ were presented in Refs.~\onlinecite{yamaguchi2004,he2006,he2006_2,yoshioka2007,varley2010,Sturm_2015,navarro2015,Ricci_2016,Furthmuller_2016}, however, few of the authors specify the coordinates of high symmetry points and lines. Lines between high symmetry points are not necessarily high symmetry lines. In particular, the coordinates  specified in previous theory works define high-symmetry points whose connecting lines do not present high symmetry lines. As pointed out by Peelaers and Van de Walle\cite{Peelaerspssb2015Ga2O3meff} most band structures published to date for $\beta$-Ga$_2$O$_3$ do not consider all high-symmetry points, and include very few high symmetry lines, thus missing significant physics of the material. In the present manuscript we are expanding the results presented by Peelaers and Van de Walle\cite{Peelaerspssb2015Ga2O3meff}, and we study the effect of the choice of the lattice vectors onto the Brillouin zone and the resulting band structure.

The dielectric tensor of $\beta$-Ga$_2$O$_3$ has been studied computationally by many authors. The most recent studies on the subject\cite{Sturm_2015,Ricci_2016,Furthmuller_2016} use the following procedure: (1) ground state calculations at the generalized gradient approximation density functional theory (GGA-DFT) level, followed by (2) calculation of the wavefunction using a hybrid functional mixing a fraction of the Hartree-Fock exchange (HF) into the GGA-DFT; (3) the hybrid wavefunction is then used for quasiparticle correction in the GW approximation\cite{hedin1965}, and (4) the dielectric tensor components including excitonic effects are calculated by solving the Bethe-Salpeter equation (BSE)\cite{BSE_1951,csanak1971}. The quality of the wavefunction obtained in step (2) has critical impact onto the quality of the resulting band structure. The first simple test of that quality is whether the chosen hybrid functional satisfactorily reproduces the experimental band gap. A recent comprehensive paper by Furthm\"uller and Bechstedt\cite{Furthmuller_2016} as well as papers by other authors\cite{varley2010,navarro2015,Sturm_2015,Ricci_2016,Furthmuller_2016} use hybrid functionals of Heyd, Scuseria, and Ernzerhof (HSE)\cite{HSE03,*HSE06} for $\beta$-Ga$_2$O$_3$. However, we observe that the HSE band gap energy converges to a value of about 4~eV, significantly less than the experimental value. We tested all popular hybrid functionals: B3LYP\cite{Becke_1993,lee1988} (published results for $\beta$-Ga$_2$O$_3$ using B3LYP functional in Ref.~\onlinecite{he2006_2}), PBE0\cite{Adamo_1999PBE0}, HSE, and Gau-PBE\cite{song2011,song2013} and found that all of them except HSE produce large band values around 4.7~eV while only HSE produces band gap value around 4~eV. Varley~\textit{et al.}\cite{varley2010} and Peelaers and Van de Walle\cite{Peelaerspssb2015Ga2O3meff} addressed this problem by increasing the fraction of direct exchange in HSE functional from 25\% to 35\% to reproduce the experimental band gap of $\beta$-Ga$_2$O$_3$. However, a most recent study by De\'ak \textit{et al.}\cite{Deak_2017} shows that increasing the amount of HF exchange compensates the effect of the screening parameter, which is the reason of the failure of the HSE functional in the case of $\beta$-Ga$_2$O$_3$. Removing the screening, on the other hand reduces the HSE functional to the PBE0 functional. In this work, we use  Gau-PBE which to the best of our knowledge has not yet been used for band structure calculation of $\beta$-Ga$_2$O$_3$. The primary purpose of our DFT calculations is to identify band-to-band transitions and to calculate parameters of CP transitions, and compare these with contributions to the experimental dielectric function of $\beta$-Ga$_2$O$_3$. For this purpose, the band structure calculations at the hybrid HF-DFT level are sufficient.\cite{paier2008,Deak_2017} Hence, we only perform steps (1) and (2) above in this present work.

Sturm~\textit{et al.}\cite{Sturm_2015} recently investigated the NIR-UV dielectric function tensor elements of $\beta$-Ga$_2$O$_3$ and noted the same observation made earlier by Jellison~\textit{et al.}, where a fourth, independent tensor element was needed to correctly model calculate  measured GSE data from multiple samples and from multiple sample azimuth orientations. Following up, Sturm~\textit{et al.}\cite{Sturm_2016} then provided the first CP analysis of monoclinic dielectric function spectra by adapting the concept of dielectric eigen displacement polarizations in the Born and Huang model towards electronic band-to-band transitions and associated exciton contributions for $\beta$-Ga$_2$O$_3$. In their analysis, Sturm~\textit{et al.} described a dipole analysis of the dielectric function by including Tanguy's model into their direction dependent  MDF approach and by assuming a band-to-band transition independent exciton binding energy parameter. In this present work, we also adapt the Born and Huang model which we have described for use in the Far-IR-IR spectral region for accurate analysis of GSE data of $\beta$-Ga$_2$O$_3$ prior to Sturm~\textit{et al.}'s work \cite{Schubert_2016}. In the eigen polarization approach, transitions are associated with a displacement direction dependent transition amplitude in analogy to the quantum mechanical concept where the transition matrix element is proportional to the transition probability depending on the choice of coordinates. The coordinates describe a given electronic system relative to the polarization direction of the incident photon. Instead of using Tanguy's expression for bound and unbound Wannier exciton contributions, we use functions described by Higginbotham, Cardona and Pollak (HCP)\cite{HigginbottamPR1969} for contributions from 3-dimensional Van-Hoofe singularities, and we augment asymmetrically broadened Lorentz oscillators to account for excitonic contributions. As a result, we find direction dependent (band-to-band transition dependent) exciton binding energies as well as band-to-band transition energies for $\beta$-Ga$_2$O$_3$ which differ in detail from those reported by Sturm~\textit{et al.}, and we provide additional information on band-to-band transitions into the VUV range not previously reported. We compare our observations with effective mass parameters determined in our DFT analysis for the topmost valence and lowest conduction bands, and we discuss recent findings reported in the literature.

\section{Theory}

\subsection{Mueller matrix generalized ellipsometry}
Ellipsometry is an indirect measurement technique that utilizes changes in the polarization state of light transmitted through or reflected off a sample surface. In general, optically anisotropic materials necessitate the application of generalized ellipsometry\cite{Schubert_1996}. We have previously provided detailed discussion on a variety of single crystalline materials and thin films with uniaxial and biaxial anisotropy using this technique\cite{Schubert_2004_2,Schubert_2002,Tiwald_2000,Schubert_1997,Schubert_1996,Dressel_2008}. Here, we investigate off-axis cut surfaces of $\beta$-Ga$_2$O$_3$ single crystals. Multiple samples cut at different angles from the same crystal are investigated using Mueller matrix generalized ellipsometry (MMGE) at multiple angles of incidence and multiple sample azimuthal angles, and all data are then analyzed simultaneously.

The Mueller matrix connects incident and emergent real-valued Stokes vector components by:
\begin{equation}
\begin{pmatrix} S_0\\ S_1\\ S_2\\ S_3\\ \end{pmatrix}_{output}= \begin{pmatrix} M_{11} & M_{12} & M_{13} & M_{14}\\ M_{21} & M_{22} & M_{23} & M_{24}\\ M_{31} & M_{32} & M_{33} & M_{34}\\ M_{41} & M_{42} & M_{43} & M_{44} \end{pmatrix}\begin{pmatrix} S_0\\ S_1\\ S_2\\ S_3\\ \end{pmatrix}_{input},
\label{MM}
\end{equation}
where the Stokes vectors are described as:
\begin{equation}
\begin{pmatrix} S_0\\ S_1\\ S_2\\ S_3\\ \end{pmatrix} = \begin{pmatrix} I_p+I_s\\ I_p-I_s\\ I_{+45}-I_{-45}\\ I_{\sigma+}-I_{\sigma-}\\ \end{pmatrix},
\end{equation}
with $I_p$, $I_s$, $I_{+45}$, $I_{-45}$, $I_{\sigma+}$, and $I_{\sigma-}$ refer to intensities with \textit{p}, \textit{s}, +45$^\circ$, -45$^\circ$, right-hand, and left-hand polarization, respectively\cite{Fujiwara_2007}.

In order to extract physical parameters from a GSE measurement (MMGE data), appropriate physical boundary conditions must be applied. For this investigation we use the so-called substrate-ambient approximation, where the single crystalline $\beta$-Ga$_2$O$_3$ samples correspond to the half-infinite substrate\cite{Fujiwara_2007}. For the substrate, the only unknown physical parameters are the dielectric function tensor elements\cite{Schubert1996}. For monoclinic materials, in order to obtain physically meaningful dielectric function tensor elements, it is necessary to assign coordinate relations between laboratory coordinate axes ($\hat{x}$,$\hat{y}$,$\hat{z}$) and crystallographic axes (\textbf{a}, \textbf{b}, \textbf{c}).\cite{Schubert_2016} We choose the $\hat{z}$ axis of the laboratory coordinate system to be normal to the sample surface, thereby defining the sample surface as the laboratory $\hat{x}$-$\hat{y}$ plane. By our choice, the (\textit{x,y,z}) system is described in Fig.\ref{unitcell} with respect to the crystal structure. Euler angles ($\phi$, $\theta$, and $\psi$) are then determined to describe angular rotations necessary to relate (\textit{x,y,z}) with ($\hat{x}$,$\hat{y}$,$\hat{z}$). Effects of finite roughness always present on the nanoscale when measuring polished crystal surfaces must be accounted for. An effective medium approximation (EMA) approach is commonly used to mimic the effect of a very thin effective layer with thickness much smaller than all  wavelengths for data analysis\cite{Aspnes_1979}. Rigorous treatment of the combination of roughness and anisotropy has not been investigated yet, hence, an isotropic averaging approach was employed here. Thus, our roughness layer model was calculated by averaging all four dielectric tensor elements and then added together in the EMA approach assuming 50\% void.

\subsection{Monoclinic dielectric tensor description}\label{Eigensection2}

With the coordinate choices in Fig.~\ref{unitcell} we select the dielectric tensor cross-term element $\varepsilon_{xy}$ as the fourth independent complex dielectric function tensor element to describe the monoclinic properties in the corresponding \textbf{a-c} plane, and the remaining elements we set to zero
\begin{equation}
\varepsilon = \begin{pmatrix} \varepsilon_{xx} & \varepsilon_{xy} & 0\\ \varepsilon_{xy} & \varepsilon_{yy} & 0\\ 0 & 0 & \varepsilon_{zz}\end{pmatrix}.
\end{equation}

\subsection{Eigen polarizability critical point model}\label{Eigensection}
We adopt the concept of the Born and Huang model and consider electronic contributions to the dielectric response of monoclinic $\beta$-Ga$_2$O$_3$ as the result of eigen dielectric displacement processes. Each individual contribution $l$ is characterized by a CP model function, $\varrho_l(\omega)$ and its eigen dielectric polarizability unit vector, $\hat{e}_l$. The same approach was adopted by Schubert~\textit{et al.}\cite{Schubert_2016} and Sturm~\textit{et al.}\cite{Sturm_2016} for analysis of GSE data for FIR-IR and NIR-VUV spectral regions, respectively, as follows:
\begin{equation}
\varepsilon(\omega) = \sum_{l=0}^{N}\varrho_l(\omega)(\hat{e_l}\otimes\hat{e_l}). \label{eigen}
\end{equation}
When \textit{l}=0, the quasi-static ($\omega \rightarrow 0$) dielectric tensor dyadic $\hat{\varepsilon}_{\mathrm{DC}}$ is determined. For monoclinic materials the tensor can be described according to:
\begin{subequations}
\begin{eqnarray}
\varepsilon_{xx}& = &\varepsilon_{\mathrm{DC},xx} + \sum_{j=1}^{m}\varrho_{j}\text{cos}^2\alpha_j ,\\
\varepsilon_{yy}& = &\varepsilon_{\mathrm{DC},yy} + \sum_{j=1}^{m}\varrho_{j}\text{sin}^2\alpha_j ,\\
\varepsilon_{xy}& = &\varepsilon_{\mathrm{DC},xy} + \sum_{j=1}^{m}\varrho_{j}\text{cos}\alpha_j\text{sin}\alpha_j ,\\
\varepsilon_{zz}& = &\varepsilon_{\mathrm{DC},zz} + \sum_{k=1}^{n}\varrho_{k} ,\\
\varepsilon_{xz}& = &\varepsilon_{yz} = 0 ,
\end{eqnarray}
\label{eq:alleps}
\end{subequations}
with $\alpha$$_j$ equal to the angle of the shear projection into the \textbf{a-c} plane and $m$,$n$ equal to the number of CP contributions in the \textbf{a-c} plane and \textbf{b} direction, respectively.

\paragraph{Fundamental band-to-band transitions:} We use photon energy ($\hbar \omega$) dependent functions described by Higginbotham, Cardona and Pollak\cite{HigginbottamPR1969} for rendering electronic contributions at 3-dimensional Van-Hoofe singularities (``M$_0$''-type CP in Adachi's CP composite approach\cite{AdachiJAPCdTe1993}):
\small
\begin{equation}
\varepsilon(E) = AE^{-1.5}\{\chi^{-2}{[2-{(1+\chi)}^{0.5}-{(1-\chi)}^{0.5}]}\}, \label{cpm0}
\end{equation}
\normalsize
with $\chi$ = ($\hbar \omega+iB)/E$, and $A$, $E$, and $B$ are, respectively, amplitude, transition energy, and broadening parameters, and $i^2=-1$. Our choice is directed by inspection of the symmetry and band curvatures for the lowest band-to-band transitions observed in our DFT calculations.

\paragraph{Exciton contributions at fundamental band-to-band transitions:} The contributions to the dielectric function due to exciton absorption arise from two parts, one from bound states and another from continuum states\cite{Yubook2005,Klingshirnbook,GrundmannPoSbook}. For Wannier-type excitons, Tanguy developed model functions for parabolic bands taking into account bound and unbound states\cite{Tanguy_1995,TangyuPRL1996,Tanguy_1999}. These functions, strictly valid for parabolic bands and isotropic materials only, were used by Sturm~\textit{et al.}\cite{Sturm_2016} for analysis of GSE data from $\beta$-Ga$_2$O$_3$. In their work, Sturm~\textit{et al.} did not determine the exciton binding energy parameter from using the Tangyu model approach. In our present work, and because the dominant contribution to exciton absorption processes in direct-band gap semiconductors is the ground state ($n$=1) transition\cite{Yubook2005,Klingshirnbook}, we employ a single Lorentz oscillator with non-symmetric broadening to account for, and to spectrally locate the ground state excitonic contribution, thereby further following Adachi's CP composite approach\cite{AdachiJAPCdTe1993}

\begin{equation}
\varepsilon = \frac{A^2-i b \hbar \omega}{E^2-(\hbar \omega)^2-i B \hbar \omega},
\label{alorentz}
\end{equation}

\noindent with $A$, $E$, $B$, and $b$ are, respectively, amplitude, energy, broadening, and asymmetric broadening parameter, respectively. The exciton binding energy parameter, $R^{\star}$[eV] can then be approximated by

\begin{equation}
\label{eq:excitonbinding}
R^{\star}_{\mathrm{y}}\approx E_{t}-E_{x}=13.6eV \left(\varepsilon^{-1}_{\infty} \right)_{jj}\left(\mu^{\star -1}\right)_{jj},
\end{equation}

\noindent where $E_t$ and $E_x$ are the band-to-band transition energy and the energy of the respective exciton CP contribution, respectively, $\left(\varepsilon^{-1}_{\infty} \right)_{jj}$ is the element $jj$ of the inverse dielectric function tensor at photon energies above the phonon mode spectral region, and $\left(\mu^{\star -1}\right)_{jj}$ is the element $jj$ of the inverse effective reduced mass tensor, for any given unit vector direction $\hat{\mathbf{j}}$. The inverse effective reduced mass tensor element is obtained from summation over the inverse effective mass tensor element of conduction and valence bands participating in the transition: $\left(\mu^{\star -1}\right)_{jj}=\left(m^{\star -1}_{cb}\right)_{jj}+\left(m^{\star -1}_{vb}\right)_{jj}$, i.e., electron and hole effective mass parameters along polarization direction $\hat{\mathbf{j}}$, respectively.
\paragraph{Above-band gap band-to-band transitions:}
At photon energies far above the band gap, multiple transitions originating at multiple points in the Brillouin zone often overlap, and  CP features due to individual transitions cannot be differentiated by experiment. Hence, broadened Lorentzian or Gaussian oscillators are often used to account for broad CP features typical for above-band gap spectra. Here, we use the same anharmonic broadened functions as in Eq.~\ref{alorentz}.

\paragraph{Higher energy band-to-band transitions:} Transitions above the spectral range investigated here contribute to the overall lineshape of the dielectric functions at wavelengths within the investigated spectral range. Such higher energy contributions are usually accounted for by Gaussian broadened oscillator functions:
\begin{eqnarray}
\varepsilon_2(\hbar \omega)& = &A(e^{-(\frac{\hbar \omega-E}{\sigma})^2}-e^{-(\frac{\hbar \omega+E}{\sigma})^2}). \label{e2} \\
\sigma& = &\frac{B}{2\sqrt{ln(2)}},  \nonumber
\end{eqnarray}
where amplitude $A$, center energy $E$, and broadening $B$ are adjustable parameters. The real part, $\varepsilon_1$, is obtained by the Kramers-Kronig integration\cite{Meneses2006}:
\begin{equation}
\varepsilon_1(\zeta) = \frac{2}{\pi}P\int_0^\infty \frac{\xi\varepsilon_2(\xi)}{\xi^2-\zeta^2}d\xi.
\end{equation}
Note that each non-trivial sum in Eqs.~\ref{eq:alleps} satisfies the Kramers-Kronig integral condition\cite{Dressel_2002,Schubert_2016,SchubertPRL2016}, and which can be set as additional side condition during the CP-MDF analysis.

\subsection{Density Functional Theory}
\begin{figure}
\includegraphics[width=8cm]{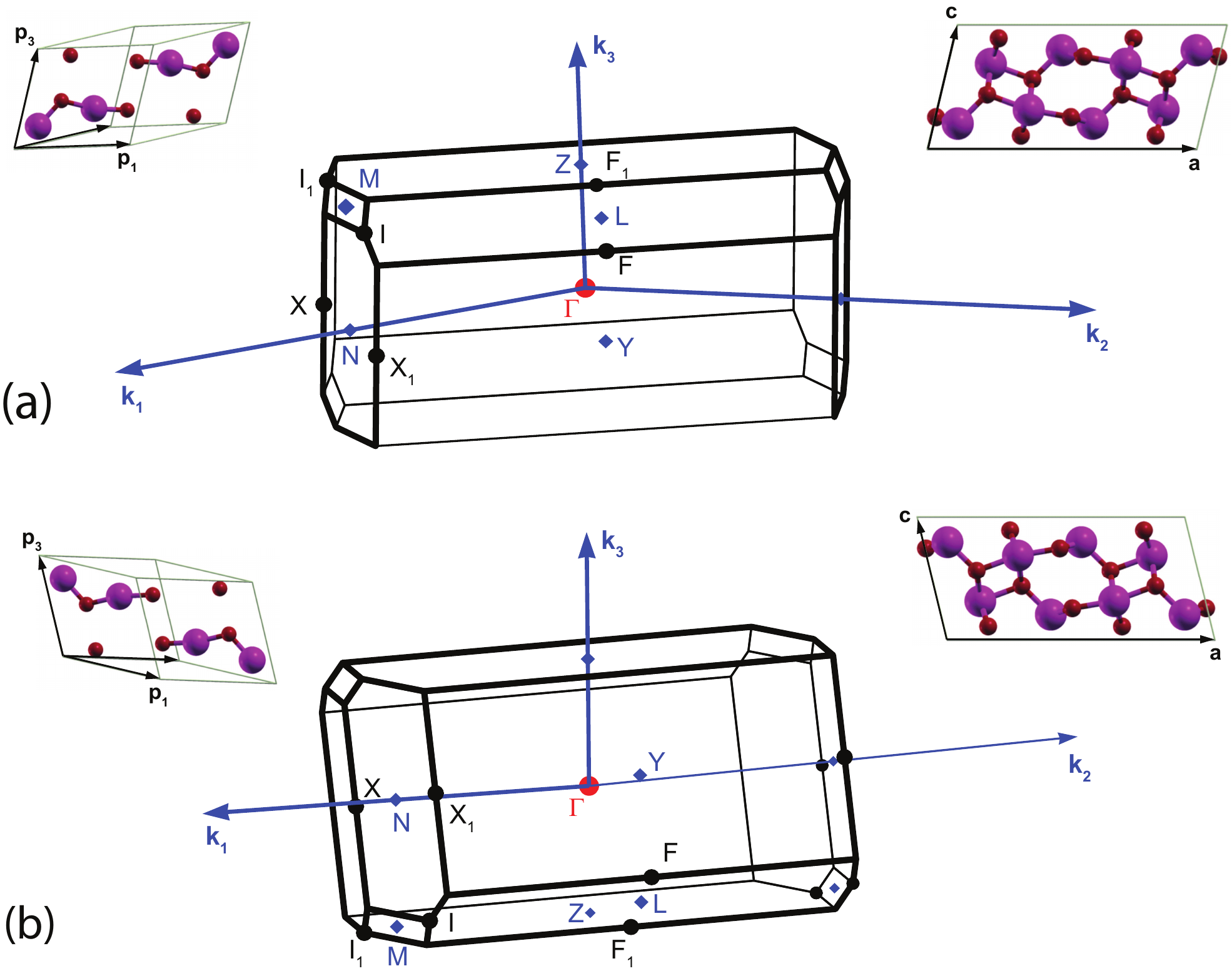}
\caption{\label{fig:bz} (Color online) Brillouin zones and alternative definitions of crystal cells for $\beta$-Ga$_2$O$_3$. (a) The Brillouin zone for primitive cell with $\beta<90^{\circ}$, with marked high symmetry points as in Ref.~\onlinecite{setyawan2010} (ibrav=-13 in Quantum ESPRESSO\cite{GiannozziJPCM2009QE}); (b) Brillouin zone for primitive cell with $\beta>90^{\circ}$ used in the present study; Primitive cells used to generate the Brillouin zones are shown to the left, and the corresponding conventional cells projected along the crystallographic axis \textbf{b} to the right.}
\end{figure}

Theoretical calculations were performed using plane wave density functional theory (DFT) code Quantum ESPRESSO (QE)\cite{[{Quantum ESPRESSO is available from http://www.qu\-an\-tum-es\-pres\-so.org. See also: }]GiannozziJPCM2009QE}. The primitive cell of $\beta$-Ga$_2$O$_3$ consisting of six oxygen and four gallium atoms was used, and the initial atomic positions and parameters of the unit cell were taken from Ref. \onlinecite{ahman1996}. The atoms were represented by norm-conserving pseudopotentials from the QE library. Following our previous calculation of phonon modes in $\beta$-Ga$_2$O$_3$\cite{Schubert_2016}, we do not include the gallium 3\textit{d} electrons in the valence configuration. Wright and Nelson\cite{wright1994} in their computational studies for GaN observed a significant overlap and hybridization of gallium 3\textit{d} states with 2\textit{s} energy levels of nitrogen, and suggested explicit treatment of these shallow states for gallium compounds. However, band structures of $\beta$-Ga$_2$O$_3$ published to date (for example Refs. \onlinecite{he2006,Furthmuller_2016}) clearly show that the gallium 3\textit{d} levels are positioned not only far ($>$10~eV) below the valence band maximum, but also well separated from other states, and little hybridization can be observed within the energy window considered in the current study. Therefore we opted for pseudopotentials for gallium that leave the 3\textit{d} electrons in the core.

First, the structure was carefully relaxed to force levels less than 10$^{-5}$ Ry/bohr using the exchange-correlation functional of Perdew, Burke and Ernzerhof (PBE).\cite{PBE} A 4$\times$4$\times$4 regular shifted Monkhorst-Pack grid was used for sampling of the Brillouin zone\cite{monhorst1976}. A convergence threshold of 10$^{-12}$ was used to reach self-consistency with a large electronic wavefunction cutoff of 100~Ry. The resulting lattice parameters obtained are shown in Tab.~\ref{tab:lattice} in comparison to results from previously reported studies using GGA-DFT methods. We find very good agreement between our values and those reported previously. Second, the structure fully relaxed at the PBE level was used for electronic structure calculations employing the hybrid Gau-PBE functional. This calculation was performed with a 6$\times$6$\times$6 $\Gamma$-centered Monkhorst-Pack grid (and after testing the convergence with respect to the grid of $k$-points up to 8$\times$8$\times$8), and with otherwise the same parameters as for the preceding PBE calculations. The converged Gau-PBE wavefunction was used to analyze the band structure.

\begin{table}
\caption{\label{tab:lattice} Comparison between the experimental and theoretical lattice constants (in \AA).}
\begin{ruledtabular}
\begin{tabular}{lccccccc}
&Exp.$^\textrm{a}$&Exp.$^\textrm{b}$&Calc.$^\textrm{c}$&Calc.$^\textrm{d}$&Calc.$^\textrm{e}$&Calc.$^\textrm{f}$&Calc.$^\textrm{g}$\\
\hline
$a$&12.214&12.233&12.287&12.27&12.31&12.438&12.289\\
$b$&3.0371&3.038&3.0564&3.03&3.08&3.084&3.0471\\
$c$&5.7981&5.807&5.823&5.80&5.89&5.877&5.8113\\
$\beta$&103.83&103.82&103.73&103.7&103.9&103.71&103.77\\
\end{tabular}
\end{ruledtabular}
\begin{flushleft}
\footnotesize{$^\textrm{a}${Ref.~\onlinecite{ahman1996}.}}\\
\footnotesize{$^\textrm{b}${Ref.~\onlinecite{lipinska2008}.}}\\
\footnotesize{$^\textrm{c}${This work, PBE.}}\\
\footnotesize{$^\textrm{d}${B88(exchange)+PW(correlation), Ref.~\onlinecite{he2006}.}}\\
\footnotesize{$^\textrm{e}${PBE, Ref.~\onlinecite{liu2007}.}}\\
\footnotesize{$^\textrm{f}${PBE, Ref.~\onlinecite{yoshioka2007}.}}\\
\footnotesize{$^\textrm{g}${AM05, Ref.~\onlinecite{Furthmuller_2016}.}}\\
\end{flushleft}
\end{table}

\begin{table*}
\caption{\label{tab:bz_points} Example coordinates of the high symmetry points in the Brillouin zone. Note that one can draw four symmetry-equivalent paths through each of the Brillouin zones, i.e., one for each possible irreducible BZ.}
\begin{ruledtabular}
\begin{tabular}{lcc}
Label&Coordinates&Coordinates\\
&for BZ in Fig. \ref{fig:bz}(a)&for BZ in Fig. \ref{fig:bz}(b)\\
\hline
$\Gamma$&[0,0,0]&[0,0,0]\\
Y&[1/2,1/2,0]&[1/2,1/2,0]\\
F&[1-$\zeta$,1-$\zeta$,1-$\eta$]&[1-$\zeta$,1-$\zeta$,$\eta$-1]\\
L&[1/2,1/2,1/2]&[1/2,1/2,-1/2]\\
I&[$\phi$,1-$\phi$,1/2]&[$\phi$,1-$\phi$,-1/2]\\
I$_1$&[1-$\phi$,$\phi$-1,1/2]&[1-$\phi$,$\phi$-1,-1/2]\\
Z&[0,0,1/2]&[0,0,-1/2]\\
F$_1$&[$\zeta$,$\zeta$,$\eta$]&[$\zeta$,$\zeta$,-$\eta$]\\
X$_1$&[$\psi$,1-$\psi$,0]&[$\psi$,1-$\psi$,0]\\
X&[1-$\psi$,$\psi$-1,0]&[1-$\psi$,$\psi$-1,0]\\
N&[1/2,0,0]&[1/2,0,0]\\
M&[1/2,0,1/2]&[1/2,0,-1/2]\\
Variables:&$\zeta = [2-(a/c) \cos(\beta)]/[4 \sin^2(\beta)]=0.39715$&$\zeta = [2+(a/c) \cos(\beta)]/[4 \sin^2(\beta)]=0.39715$\\
&$\eta=1/2+2 \zeta (c/a) \cos(\beta) = 0.58937$&$\eta=1/2-2 \zeta (c/a) \cos(\beta) = 0.58937$\\
&$\psi = 3/4-b^2/[4a^2 \sin^2(\beta)]=0.7336$&same\\
&$\phi = \psi+(3/4-\psi) (a/c) \cos(\beta)=0.74181$&$\phi = \psi-(3/4-\psi) (a/c) \cos(\beta)=0.74181$\\
\end{tabular}
\end{ruledtabular}
\end{table*}

The visualization program XCrySDen\cite{[][{. Code available from http://www.xcrysden.org.}]KokaljCMS2003XCrysDen} produces the proper Brillouin zones for all known sets of lattice vectors, including base-centered monoclinic, using the Voronoi decomposition. The orientation of the Brillouin zones for base center monoclinic lattices, however, depends on the definition of the primitive cell used. While commonly the symmetry axes is assigned with 'unique axis $b$' and the monoclinic angle $\beta>90^{\circ}$, Setyawan and Curtarolo\cite{setyawan2010} used an unconventional definition using 'unique axis $a$' and monoclinic angle $\alpha<90^{\circ}$. In the standard convention of using 'unique axis $b$' and the monoclinic angle $\beta>90^{\circ}$, a primitive cell with vectors $p_1 = (a+b)/2$ and $p_2 = (-a+b)/2$ produces a Brillouin zone like in Fig.~2 in Ref.~\onlinecite{Peelaerspssb2015Ga2O3meff}, whereas a primitive cell with vectors $p_1 = (a-b)/2$ and $p_2 = (a+b)/2$ produces a Brillouin zone identical with Fig.~17 in Ref.~\onlinecite{setyawan2010} if $\beta<90^{\circ}$, and one other when flipped vertically for $\beta>90^{\circ}$, shown in Fig.~\ref{fig:bz}(a) and Fig.~\ref{fig:bz}(b), respectively. In our present study, we use the structure shown in Fig.~\ref{fig:bz}(b).  If high symmetry points are correctly identified, then physical properties descending from such high symmetry points should not depend on the choice of coordinates or convention for the Brillouin zone description. For convenience, we define these points in Fig.~\ref{fig:bz} for the Brillouin zone oriented exactly like in the paper of Setyawan and Curtarolo. Our coordinates of the high symmetry points in both orientations are given in Tab.~\ref{tab:bz_points}.

The band structure along a high symmetry path was plotted using the band interpolation method based on the maximally localized Wannier functions\cite{Marzari1997,Souza2001} as implemented in the software package WANNIER90\cite{mostofi2008}. We used s and p orbitals on both Ga and O atoms and performed disentanglement of the bands in a frozen energy window from -5~eV to 22~eV. The disentangled bands were also used for calculating the effective masses of the carriers. The bands were sampled in the range $\pm$0.005~\AA$^{-1}$ from the $\Gamma$ point parallel to the crystal directions \textbf{a}, \textbf{b}, and \textbf{c}. Parabolic curves were used to fit the dispersions of the respective energy bands and the quadratic terms of the parabolas were converted to inverse effective mass tensor parameters as follows:

\begin{equation}
\left(m^{\star -1}\right)_{jj} = \frac{1}{\hbar^2} \frac{\partial^2}{\partial k_j^2}E(\bf{k}),
\end{equation}

\noindent where derivatives are taken along directions $\mathbf{k}=k_j\hat{\mathbf{j}}$ with unit vector $\hat{\mathbf{j}}$, for example, parallel to $\mathbf{a}$, $\mathbf{b}$, or $\mathbf{c}$.

Finally, the significant band-to-band transitions contributing to the experimental dielectric tensor measured by ellipsometry were identified by analyzing the matrix elements $|\mathcal{M}_{cv}|^2$ of the momentum operator between conduction and valence bands at the $\Gamma$ point (Tab.~\ref{tab:transitions}). The signatures (parallel or anti-parallel) of the projections of $|\mathcal{M}_{cv}|^2$ along the crystal directions \textbf{a} and \textbf{c$^*$} were obtained from inspecting the complex  argument of $\mathcal{M}_{cv}$. Transition eigenvectors with  parallel arguments (antiparallel) were plotted in the first (second) quadrant of the Cartesian ($\mathbf{a-c^{\star}}$) plane.

\section{Experiment}

Bulk single crystalline $\beta$-Ga$_2$O$_3$ was grown by Tamura Corp., Japan by the edge-defined film fed growth process as described in Refs.~\onlinecite{Aida_2008,Sasaki_2012,Shimamura_2013}. Samples were then cut at different orientations to the dimensions of 650~$\mu$m $\times$ 10~mm $\times$ 10~mm, and then polished on one side. In this paper we investigate a (010) and a ($\overline{2}01$) surface.

Mueller matrix generalized spectroscopic ellipsometry data were collected from 133 nm to 1690 nm. The vacuum ultraviolet (VUV) measurements were obtained using a rotating-analyzer ellipsometer with an automated compensator function (VUV-VASE, J.A. Woollam Co., Inc.). Data were acquired at three angles of incidence ($\Phi_a$=50$^\circ$, 60$^\circ$, 70$^\circ$), and at several azimuthal angles by manually rotating the sample about the sample normal in steps of $\approx $45$^\circ$. Note that in the VUV range, due to limitations of the instrument, no elements in the 4$^{th}$ row of the Mueller matrix are available. Measurements from the near infrared to near ultraviolet (NIR-NUV) were performed using a dual-rotating compensator ellipsometer (RC2, J.A. Woollam Co., Inc.) allowing for the determination of the complete 4$\times$4 Mueller matrix. Measurements were taken at three angles of incidence ($\Phi_a$=50$^\circ$, 60$^\circ$, 70$^\circ$), and at different orientations by auto-rotating the sample by steps of 15$^\circ$ beginning at the same azimuthal orientation as in the VUV measurements. All model calculations were conducted using WVASE32$^{\mathrm{TM}}$ (J.~A.~Woollam Co., Inc.).

\section{Results and Discussion}

\subsection{Wavelength-by-wavelength analysis of the dielectric function tensor }
\begin{figure*}[hbt]
    \includegraphics[width=.99\linewidth]{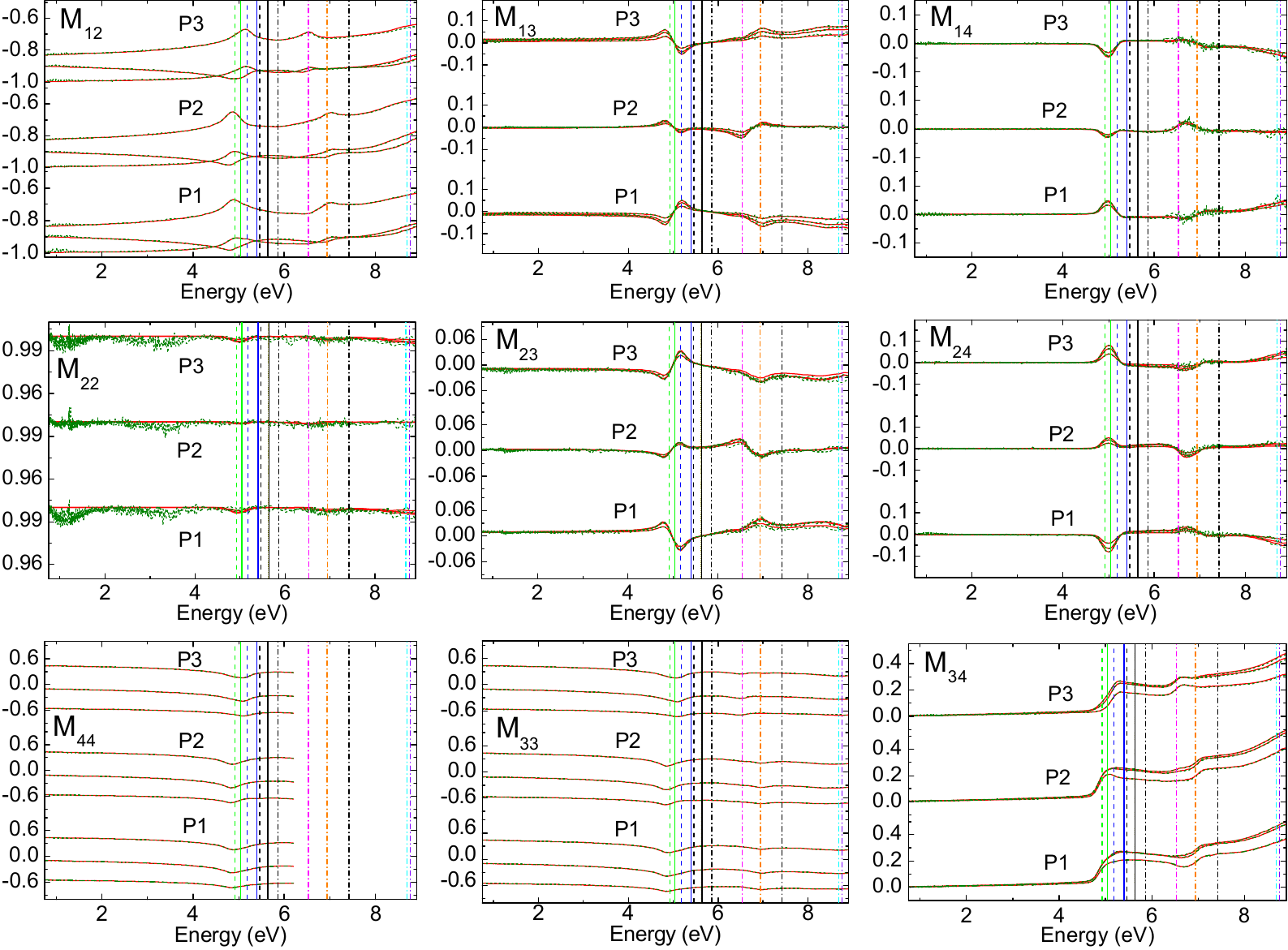}
    \caption{Experimental (dotted, green lines) and best match model (solid, red lines) Mueller matrix data obtained from $\beta$-Ga$_2$O$_3$ (010) surface at three different azimuthal orientations (P1: $\varphi$ = 38.5(1)$^\circ$, P2: $\varphi$ = 77.4(1)$^\circ$, P3: $\varphi$ = 130.42(1)$^\circ$). Data were taken at three angles of incidence ($\Phi_a$ = 50$^\circ$, 60$^\circ$, 70$^\circ$). Vertical lines indicate energies at which CP transitions were suggested by the lineshape analysis. For color code and line styles of vertical lines, refer to Fig.~\ref{energies}. Euler angle parameters $\theta$ = -0.04(1)$^\circ$ and $\psi$ = 0.0(1)$^\circ$ are consistent with the crystallographic orientation of the (010) surface.}
    \label{010datafit}
\end{figure*}
\begin{figure*}[hbt]
    \includegraphics[width=.99\linewidth]{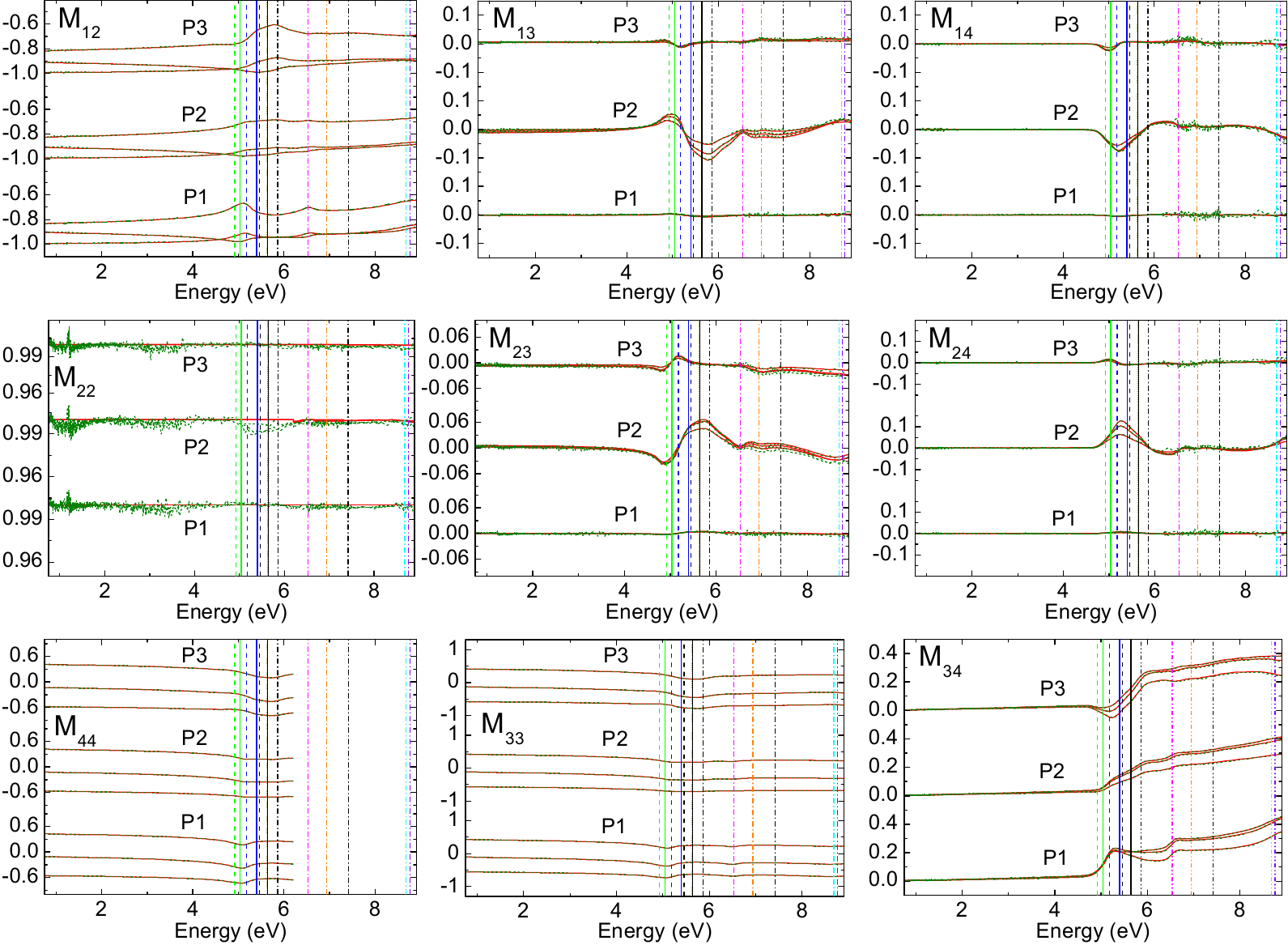}
    \caption{Same as Fig.~\ref{010datafit} except for $\beta$-Ga$_2$O$_3$ ($\overline{2}01$) surface. (P1: $\varphi$ = 184.3(1)$^\circ$, P2: $\varphi$ = 228.9(1)$^\circ$, P3: $\varphi$ = 266.7(1)$^\circ$). Euler angle parameters $\theta$ = 89.97(1)$^\circ$ and $\psi$ = -52.9(1)$^\circ$ are consistent with the crystallographic orientation of the ($\overline{2}01$) surface.}
    \label{2b01datafit}
\end{figure*}
Experimental and best match model calculated Mueller matrix GSE data is summarized in Figs.~\ref{010datafit} and~\ref{2b01datafit} for the (010) and ($\overline{2}01$) surfaces, respectively. Selected data, obtained at 3 different sample azimuthal orientations 45$^\circ$ apart, and 3 angles of incidence (50$^\circ$, 60$^\circ$ and 70$^\circ$) are displayed. Panels with individual Mueller matrix elements are shown separately and arranged according to their indices. All Mueller matrix data are normalized to element M$_{11}$. For non-magnetic and non-chiral materials, in general, and as can be seen in the experimental and calculated data, Mueller matrix elements with symmetric indices can be obtained from simple symmetry operation, thus only the upper diagonal elements are presented. Data are shown for energies 0.75-9~eV except for $M_{44}$ which only contains data from approximately 0.75-6.2~eV. Data gathered from additional azimuthal orientations are not shown.

\begin{figure}[hbt]
   \includegraphics[width=.95\linewidth]{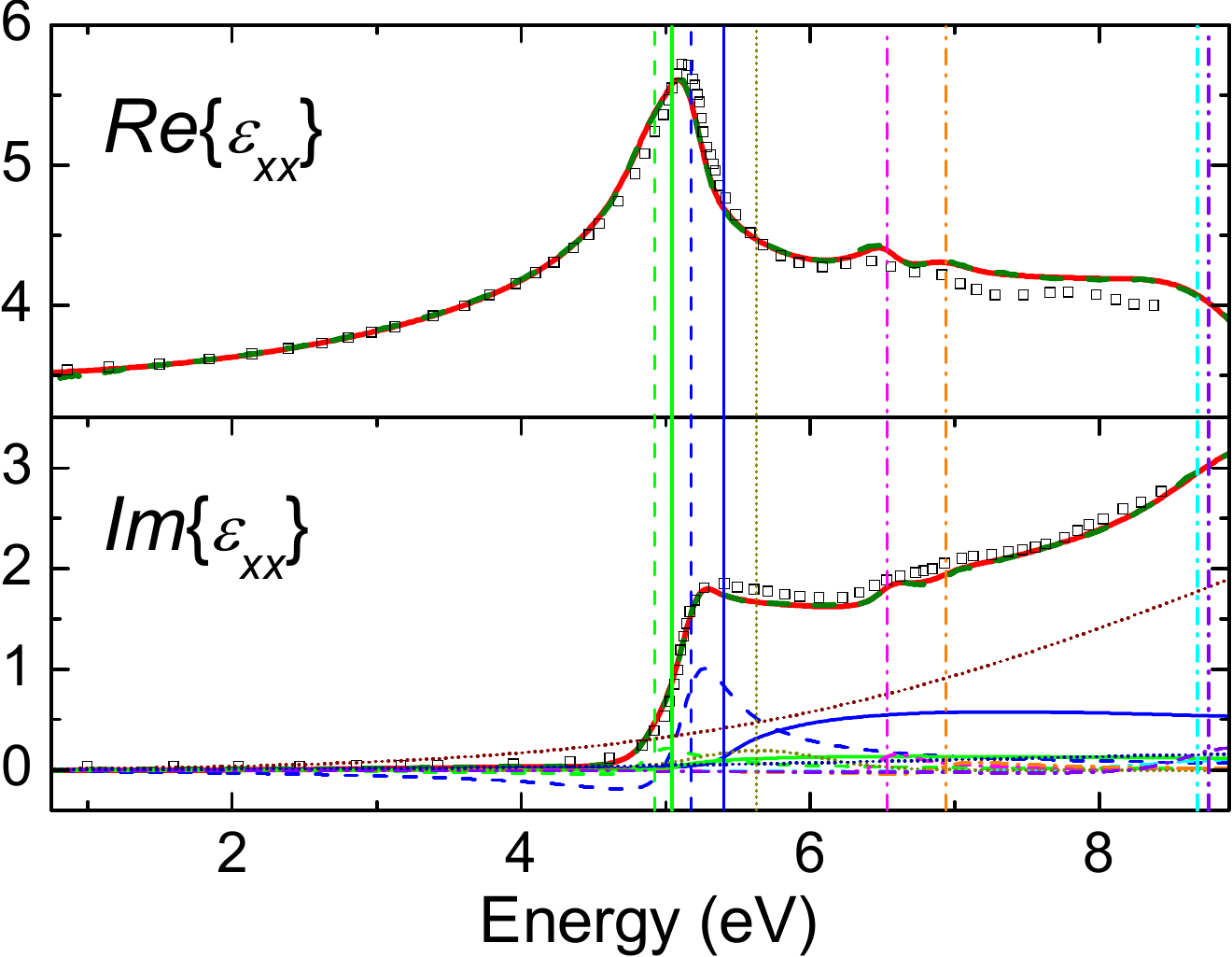}
\caption{Dielectric function tensor element $\varepsilon_{xx}$, approximately along axis \textbf{a} in our coordinate system, obtained from wavelength-by-wavelength (polyfit) analysis (green dotted lines) and best match MDF analysis (red solid lines). Individual CP contributions to the MDF are shown for the imaginary part only. Vertical lines indicate CP transition energy model parameters obtained from MDF analysis. For color code and line styles refer to Fig.~\ref{energies}. Data from MDF analysis by Sturm~\textit{et al.} are included for comparison (Ref.~\onlinecite{Sturm_2015}; open symbols). }
    \label{exx}
\end{figure}

\begin{figure}[hbt]
\includegraphics[width=.95\linewidth]{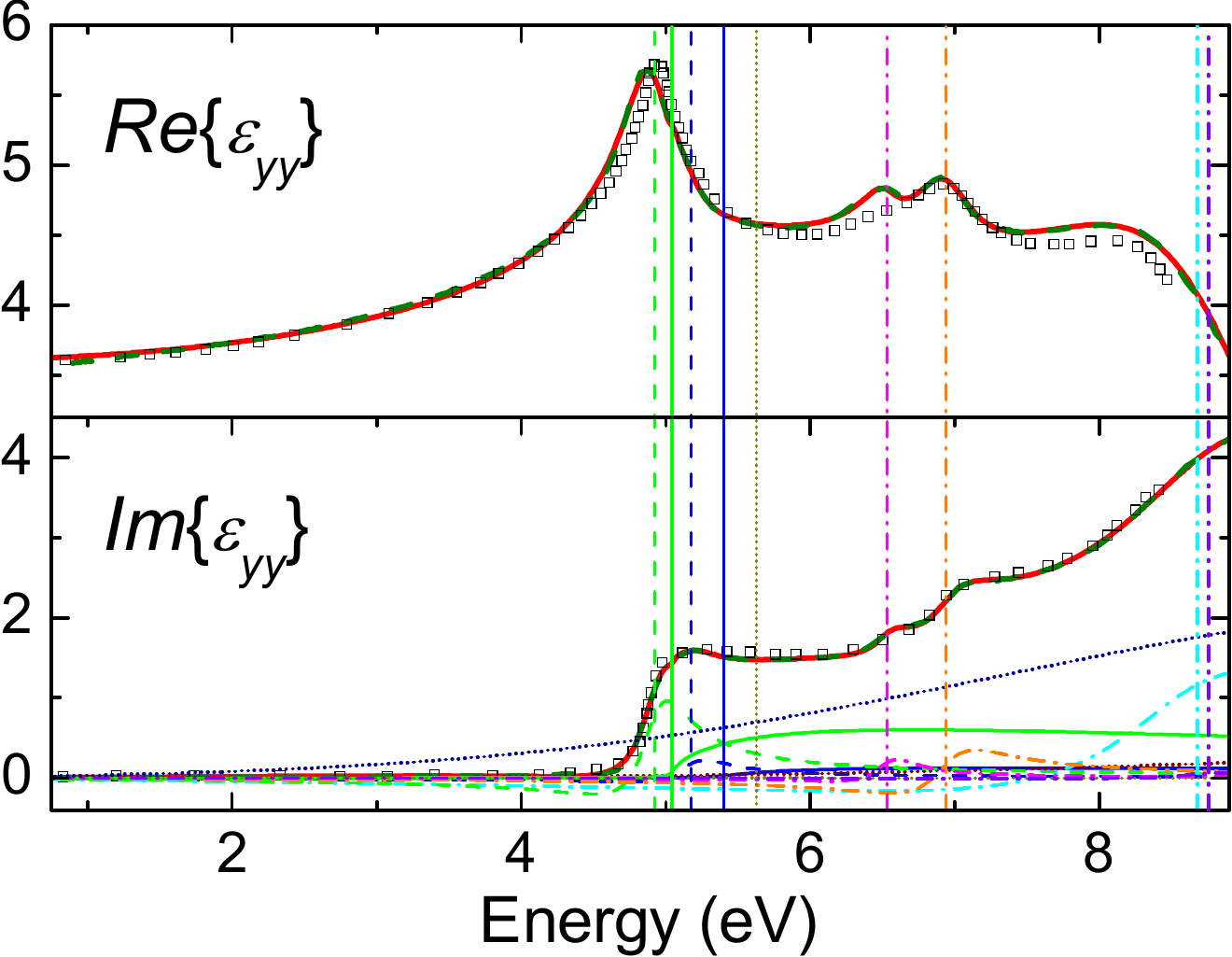}
    \caption{Same as Fig.~\ref{exx} for $\varepsilon_{yy}$ approximately along axis \textbf{c$^*$}.}
    \label{eyy}
\end{figure}

\begin{figure}[hbt] \includegraphics[width=.95\linewidth]{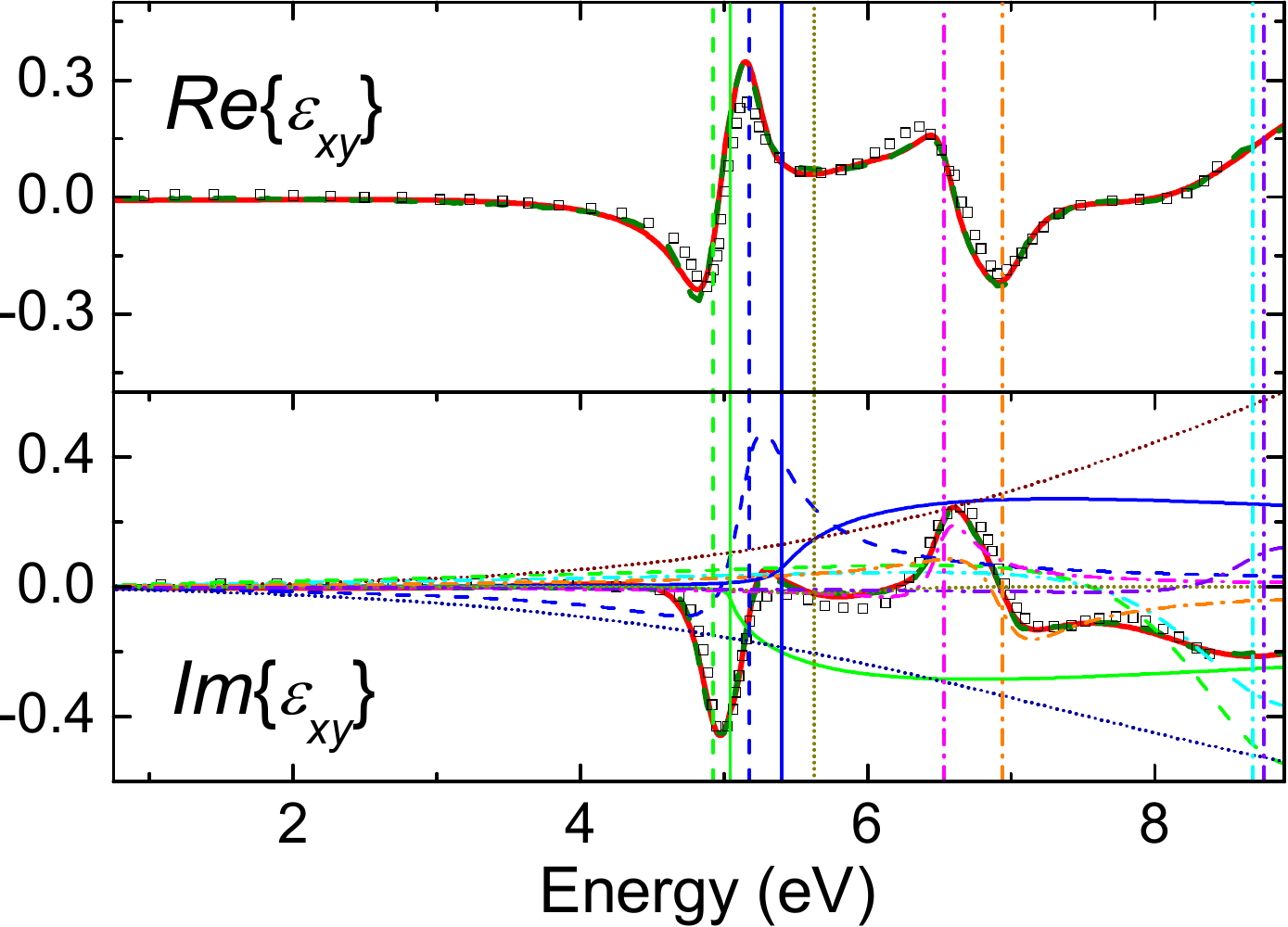}
    \caption{Same as Fig.~\ref{exx} for $\varepsilon_{xy}$ within the \textbf{a-c} plane.}
    \label{exy}
\end{figure}

\begin{figure}[hbt]
\includegraphics[width=.95\linewidth]{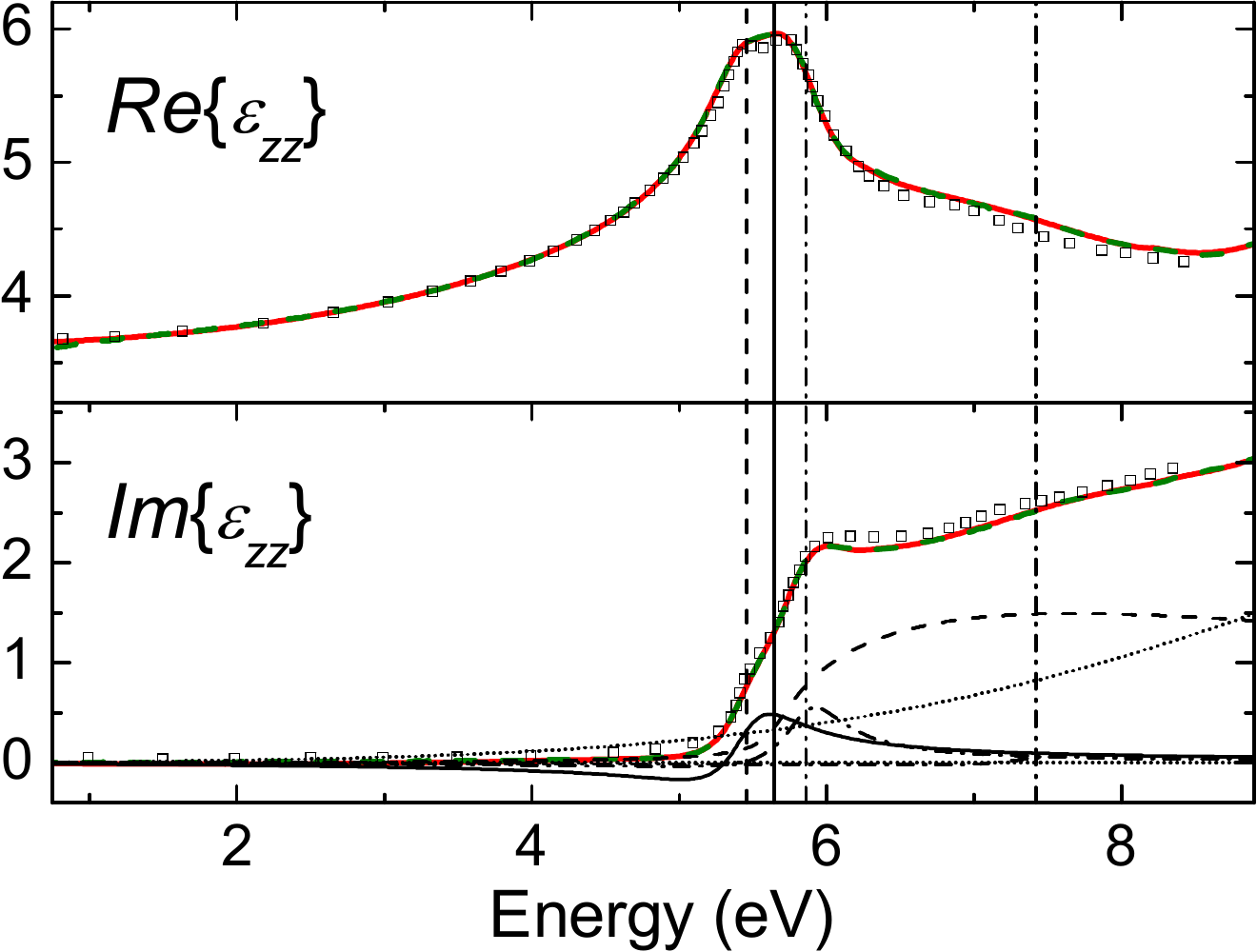}
    \caption{Same as Fig.~\ref{exx} for $\varepsilon_{zz}$ approximately along axis \textbf{b}.}
    \label{ezz}
\end{figure}

Each data set (sample, azimuthal orientation, angle of incidence) is unique, however, characteristic features are shared between them all at energies indicated by vertical lines. While we do not show all data in Figs. \ref{010datafit} and \ref{2b01datafit}, we note that all data sets are identical when samples are measured at 180$^\circ$ rotated azimuth orientation. Most important to note in the experimental Mueller matrix data is the clear anisotropy shown by the nonzero off-diagonal block elements (M$_{13}$, M$_{14}$, M$_{23}$, M$_{24}$) and strong dependence on sample azimuthal orientation in all Mueller matrix elements. All data gathered by the measurement of multiple samples, with multiple orientations, and at multiple angles of incidence were analyzed simultaneously using a best-match model data regression procedure (polyfit). For each energy, up to 144 independent data points were included from 2 samples, 3 angles of incidence, and as many as 24 different azimuthal orientations. Only 8 independent model parameters for real and imaginary parts of $\varepsilon_{xx}$, $\varepsilon_{yy}$, $\varepsilon_{zz}$, $\varepsilon_{xy}$ as well as 2 sets of energy-independent Euler angles describing the sample orientation and crystallographic structure and 2 roughness layer thickness parameters were fit for. The thickness parameters for the roughness layer of the (010) and ($\overline{2}01$) samples were determined to be 1.78(1)~nm and 1.61(1)~nm, respectively. The best match model calculated Mueller matrix elements from the polyfit procedure are shown in Figs.~\ref{010datafit} and~\ref{2b01datafit} as red solid lines. We obtain an excellent agreement between model calculated and experimental Mueller matrix data. Euler angle parameters noted in the captions of Figs.~\ref{010datafit} and~\ref{2b01datafit} are in agreement with anticipated orientations of the crystallographic axes of each of the samples. The dielectric function tensor elements, $\varepsilon_{xx}$, $\varepsilon_{yy}$, $\varepsilon_{xy}$, and $\varepsilon_{zz}$ determined from the wavelength-by-wavelength polyfit procedure are shown in Fig.~\ref{exx}, Fig.~\ref{eyy}, Fig.~\ref{exy} and Fig.~\ref{ezz}, respectively, as dotted green lines.

\subsection{CP model analysis}
 We identify 11 differentiable contributions in $\varepsilon_{xx}$, $\varepsilon_{yy}$, and $\varepsilon_{xy}$, and 5 in $\varepsilon_{zz}$. Distinct features can be seen, e.g., in the imaginary part of each tensor element in Figs.~\ref{exx}-\ref{ezz}. Vertical lines are drawn corresponding to the results from our CP analysis at the respective CP transition energy model parameters. Note that while vertical lines are identical for $\varepsilon_{xx}$, $\varepsilon_{yy}$, and $\varepsilon_{xy}$, a different set is seen for $\varepsilon_{zz}$ which corresponds to the difference between the monoclinic \textbf{a-c} plane and the axis parallel to \textbf{b}.

\paragraph{\textbf{a-c} plane} Eleven CP features are needed to match the tensor elements. Functions described in Sect.~\ref{Eigensection} are used to model individual CP contributions as projections in the $x$-$y$ plane with angular parameters $\alpha_{\mathrm{CP},j}$. The lowest band-to-band transition, CP$^{\mathrm{ac}}_{0}$ was modeled with the HCP CP function (Eq.~\ref{cpm0}) with an excitonic contribution determined by an asymmetrically broadened Lorentzian oscillator (Eq.~\ref{alorentz}). It was assumed that excitonic and band-to-band transition contributions share the same unit eigenvector (angular parameter $\alpha_{\mathrm{CP},0}$). The exciton CP contribution is labeled CP$^{\mathrm{ac}}_{0x}$. We identify a second pair of CP contributions (CP$^{\mathrm{ac}}_{1}$, CP$^{\mathrm{ac}}_{1x}$, $\alpha_{\mathrm{CP},1}$) using the same functions. A Gaussian oscillator was used to model a very small CP contribution at 5.64~eV which could not be further differentiated (CP$^{\mathrm{ac}}_{2}$, $\alpha_{\mathrm{CP},2}$). Above-band gap CP contributions (CP$^{\mathrm{ac}}_{3-6}$, $\alpha_{\mathrm{CP},3-6}$) were identified at higher photon energies, which were modeled by asymmetrically broadened Lorentzian oscillators (Eq.~\ref{alorentz}). We were unable to differentiate between excitonic and band-to-band transition contributions associated with these higher energy CPs. Contributions due to higher-energy transitions, outside the investigated spectral region, were accounted for by a Gaussian function CP with projection along $x$ (CP$^{\mathrm{ac}}_{7}$), and $y$ (CP$^{\mathrm{ac}}_{8}$). The resulting best-match CP-MDF parameters are listed in Tab.~\ref{parm}, and are shown as solid red lines in Figs.~\ref{exx}-\ref{exy}. An excellent agreement between our GSE wavelength-by-wavelength obtained and CP-MDF calculated data is noted. We also note close agreement with the GSE wavelength-by-wavelength obtained data reported by  Sturm~\textit{et al}.

\begin{table}[hbt]
\setlength{\tabcolsep}{4pt}
\caption{\label{parm}CP-MDF parameters for polarization within the \textbf{a-c} plane of $\beta$-Ga$_{2}$O$_{3}$ obtained in this work from MMGE wavelength-by-wavelength data analysis of (010) and ($\overline{2}01$) surfaces of single crystalline bulk $\beta$-Ga$_{2}$O$_{3}$.}
\begin{ruledtabular}
\begin{tabular}{cccccc}
 & $\alpha$ ($^\circ$) & $A$~(eV) & $E$~(eV) & $B$~(eV) & $b$~(eV) \\
\hline
CP$^{ac}_{0x}$& 115.1(1) & 1.35(1) & 4.92(1)$^\textrm{a}$ & 0.40(1) & 0.44(1) \\
$^\textrm{b}$CP$^{ac}_{0}$& 115.1(1) & 25.9(4) & 5.04(1) & 0.02(1) & - \\
CP$^{ac}_{1x}$& 25.2(1) & 1.50(1) & 5.17(1)$^\textrm{a}$ & 0.43(1) & 0.48(1) \\
$^\textrm{b}$CP$^{ac}_{1}$ & 25.2(1) & 28.0(5) & 5.40(1) & 0.09(1) & - \\
$^\textrm{c}$CP$^{ac}_{2}$ & 174.2(2) & 0.19(1) & 5.64(1) & 1.05(1) & -\\
CP$^{ac}_{3}$& 50.4(1) & 0.85(1) & 6.53(1) & 0.34(1) & 0.11(1) \\
CP$^{ac}_{4}$& 114.6(1) & 0.88(2) & 6.94(1) & 0.56(1) & 0.35(1) \\
CP$^{ac}_{5}$& 105.4(1) & 4.60(5) & 8.68(1) & 1.94(1) & 2.24(4) \\
CP$^{ac}_{6}$& 29.2(1) &  1.45(4) & 8.76(1) & 0.97(2) & 0.22(1) \\
$^\textrm{c}$CP$^{ac}_{7}$& 106.4(1) &  2.34(1) & 10.91(1) & 8.28(1) & - \\
$^\textrm{c}$CP$^{ac}_{8}$& 17.6(1) &  3.56(1) & 12.54(1) & 8.28(1) & - \\
\end{tabular}
\end{ruledtabular}
\begin{flushleft}
\footnotesize{$^\textrm{a}${Energy calculated from binding energy model parameter.}}\\
\footnotesize{$^\textrm{b}${Denotes 3D M$_0$ Adachi function.}}\\
\footnotesize{$^\textrm{c}${Denotes Gaussian oscillator used in this analysis.}}\\
\end{flushleft}
\end{table}

\paragraph{\textbf{b} axis} Six CP features are needed to match the dielectric tensor element $\varepsilon_{zz}$. Functions described in Sect.~\ref{Eigensection} are used to model individual CP contributions projected along axis \textbf{b}. The lowest band-to-band transition, CP$^{\mathrm{b}}_{0}$ was modeled with the HCP CP function (Eq.~\ref{cpm0}; CP$^{\mathrm{b}}_{0}$) with an excitonic contribution (CP$^{\mathrm{b}}_{0x}$) determined by an asymmetrically broadened Lorentzian oscillator (Eq.~\ref{alorentz}). Above-band CP contributions (CP$^{\mathrm{b}}_{1-2}$) were identified and modeled by functions in Eq.~\ref{alorentz}. Here again, we were unable to differentiate between excitonic and band-to-band transition contribution. Contributions due to higher-energy transitions, outside the investigated spectral region, were accounted for by a Gaussian function CP with projection along $z$ (CP$^{\mathrm{b}}_{3-4}$). The resulting best-match CP-MDF parameters are listed in Tab.~\ref{parmbaxis}, and are shown as solid red lines in Fig.~\ref{ezz}. Again, an excellent agreement between our GSE wavelength-by-wavelength obtained and CP-MDF calculated data is noted. We also note close agreement with the GSE wavelength-by-wavelength obtained data reported by  Sturm~\textit{et al}.

\begin{table}
\setlength{\tabcolsep}{6pt}
\caption{\label{parmbaxis}Same as for Tab.~\ref{parm} but for transitions polarized parallel to axis \textbf{b}.}
\begin{ruledtabular}
\begin{tabular}{ccccc}
 & $A$~(eV) & $E$~(eV) & $B$~(eV) & $b$~(eV) \\
\hline
CP$^{b}_{0x}$&  0.97(6) & 5.46(3)$^\textrm{a}$ & 0.54(1) & 0.32(1) \\
$^\textrm{b}$CP$^{b}_{0}$ &  64(2) & 5.64(1) & 0.11(1) & - \\
CP$^{b}_{1}$& 1.24(6) & 5.86(1) & 0.50(2) & 0.15(2) \\
CP$^{b}_{2}$& 0.59(7) & 7.42(3) & 0.95(5) & 0.08(1) \\
$^\textrm{c}$CP$^{b}_{3}$ & 1.37(3) & 9.53(1) & 0.47(2) & - \\
$^\textrm{c}$CP$^{b}_{4}$& 3.50(1) & 13.82(1) & 8.86(1) & - \\
\end{tabular}
\end{ruledtabular}
\begin{flushleft}
\footnotesize{$^\textrm{a}${Energy calculated from binding energy model parameter.}}\\
\footnotesize{$^\textrm{b}${Denotes 3D M$_0$ Adachi function.}}\\
\footnotesize{$^\textrm{c}${Transition outside investigated spectral region with limited sensitivity modeled with a Gaussian oscillator.}}\\
\end{flushleft}
\end{table}

\subsection{DFT analysis}
\begin{figure}
\includegraphics[width=.95\linewidth]{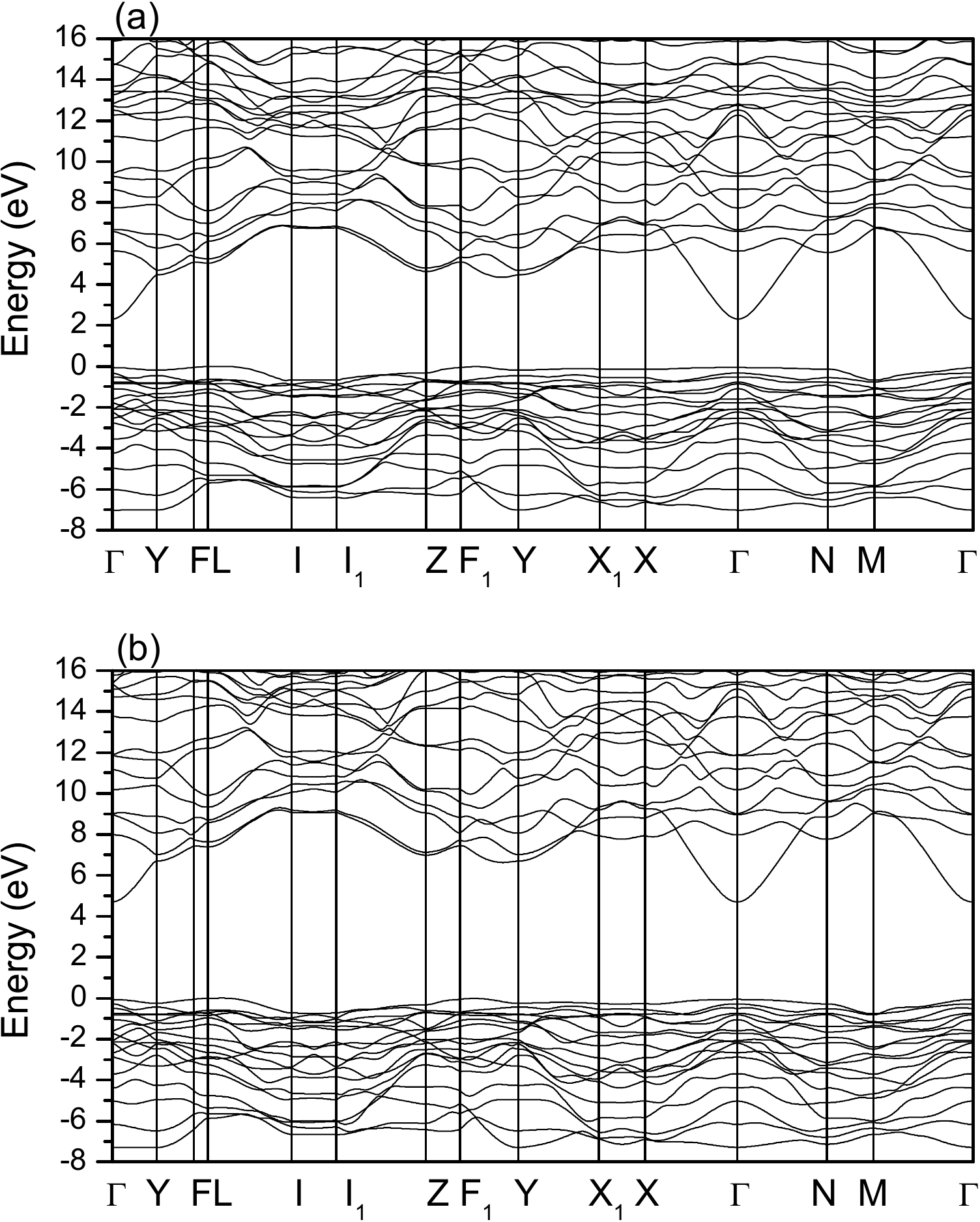}
\caption{\label{fig:bands} Band structure of $\beta$-Ga$_2$O$_3$. (a) At the GGA-DFT (PBE) level; (b) at the hybrid HF-DFT (Gau-PBE) level.}
\end{figure}
\subsubsection{Band structure}
Figure~\ref{fig:bands} shows plots of the band structure at the DFT (PBE functional) and hybrid HF-DFT (Gau-PBE functional) levels of theory. The major difference between the two plots is the expected opening of the energy gap between the valence and conduction bands by about 2~eV. The most important feature, common to both plots, is the lowest conduction band which dominates the Brillouin zone center. A direct comparison between our band structures and previously published data is rather difficult as the majority of the authors only considered some of the high symmetry points in the first Brillouin zone. The only complete band structure to date was published by Peelaers and Van de Walle\cite{Peelaerspssb2015Ga2O3meff}. They used the HSE density functional with the fraction of HF exchange adjusted to reproduce an assumed value of the band gap. Their valence bands are very similar to ours, whereas their conduction bands are slightly shifted to higher energies and steeper than ours, most likely due to the effect of the higher amount of HF exchange included into the calculations (35\% vs 24\% in Gau-PBE). Otherwise the agreement is rather good.

The character of the band gap in $\beta$-Ga$_2$O$_3$ can be obtained from the band structure. The broad valence band maximum (VBM) has been reported previously along the $L$-$I$ line of the Brillouin zone\cite{yamaguchi2004,Peelaerspssb2015Ga2O3meff}, slightly off the $L$ point. Note again that in the current manuscript we use the nomenclature and labeling proposed by Setyawan and Curtarolo\cite{setyawan2010} with point $L$=[1/2,1/2,1/2], while in most previous publications this point is labeled $M$. Due to the fact that the valence band is very flat along the $L-I$ line, the actual location of the VBM can be easily missed. However, the energy difference between the actual VBM and, for example, the top valence band energy at the $L$-point only amounts to few meV. Thus studying the band properties at high-symmetry points $L$ and at the $\Gamma$ points is accurate enough. Local density approximation DFT methods typically predict the band gap to be indirect, and render the valence band at the $L$-point about 100~meV higher than the direct gap at $\Gamma$\cite{yamaguchi2004}. At the GGA-DFT level this difference is reduced to about 20-50 meV\cite{he2006,Furthmuller_2016}, and this usually holds for hybrid HF-DFT as well\cite{he2006_2,varley2010,Peelaerspssb2015Ga2O3meff,Furthmuller_2016}. Our results at the Gau-PBE level show  the VBM near the $L$ point about 50~meV higher than the top valence band at $\Gamma$. Interestingly, at the GW level (quasiparticle bands) the band gaps are completely degenerate, or even the direct gap appears marginally higher\cite{Furthmuller_2016}. However, Ratnaparkhe and Lambrecht\cite{Ratnaparkhe_2017} used the quasiparticle self-consistent version of GW, QSGW\cite{Schilfgaarde_2006QSGW}, and obtained the indirect band gap energy smaller by nearly 100~meV than the direct band gap energy.

\subsubsection{Band-to-band transitions}
We analyze band-to-band transitions by identifying all allowed transitions, i.e., transitions with non-zero matrix elements of the momentum operator between conduction and valence bands, and whose transition energies are less than 10~eV. We find eight such transitions, summarized in Tab.~\ref{tab:transitions} and Tab.~\ref{tab:baxistransitions}, respectively, with polarization within the \textbf{a-c} plane and along axis \textbf{b}, respectively. The transition labels are according to the numbers of the bands involved, where numberings start at the top of the valence and at the bottom of the conduction bands. The matrix elements of the momentum operator are obtained from the overlap of the wavefunctions for the respective energy bands. Hence, their values represent the probabilities of the transitions, i.e., transition amplitudes, which can be compared to experimental ones. In case of \textbf{a-c} plane transitions, which do not have any pre-defined orientation within the plane, the transition probabilities along the crystallographic directions \textbf{a} and \textbf{c$^*$} constitute Cartesian components of the corresponding transition vectors, thereby defining the polarization direction of the band-to-band transition in space. Their orientations are shown in Fig.~\ref{vectors}(b), and can be compared to the eigen dielectric displacement unit vectors obtained from GSE analysis. The  fundamental (lowest energy) band-to-band transition is polarized nearly parallel to the crystallographic axis \textbf{c}. It is closely followed by a second transition polarized at a small angle from the crystallographic axis \textbf{a}. The lowest transition along the crystallographic axis \textbf{b} occurs about 0.6~eV above the fundamental transition in the \textbf{a-c} plane. This tendency agrees well with GW results shown in Tab.~VIII of Ref.~\onlinecite{Furthmuller_2016}.

\begin{table}
\setlength{\tabcolsep}{1pt}
\caption{\label{tab:transitions}Calculated band-to-band transition energies ($E$) within the \textbf{a-c} plane, and transition matrix elements $|\mathcal{M}_{cv}|^2_{a}$ and $|\mathcal{M}_{cv}|^2_{c^{\star}}$ projected onto axis \textbf{a} and $\mathbf{c^{\star}}$, respectively. Transitions are labeled $\Gamma_{c-v}$ with indexes numbering bands upwards from the bottom ($c=1$) of the conduction band and downwards from the top ($v=1$) of the valence band at the $\Gamma$ point. The polarization angle $\alpha$ is measured relative to axis \textbf{a}.}
\begin{ruledtabular}
\begin{tabular}{ccccccc}
Label&$E$ (eV)&$\alpha$($^\circ$)&$|\mathcal{M}_{cv}|_a^2$&$|\mathcal{M}_{cv}|_{c^*}^2$&$c$&$v$\\
\hline
$\Gamma_{1-1}$&4.740&100.504&0.01972523&0.10638229&1&1\\
$\Gamma_{1-2}$&4.969&7.498&0.12773652&0.01681217&1&2\\
$\Gamma_{1-7}$&6.279&74.797&0.01304504&0.0480026&1&7\\
$\Gamma_{1-11}$&6.879&129.305&0.02598545&0.03174252&1&11\\
$\Gamma_{2-3}$&8.453&34.828&0.01417296&0.00986065&2&3\\
*$\Gamma_{4-1}$&9.0163&108.953&0.00011979&0.00034883&4&1\\
*$\Gamma_{4-2}$&9.2456&75.6222&0.00075006&0.00292599&4&2\\
$\Gamma_{3-3}$&9.432&88.912&0.0006232&0.03281954&3&3\\
$\Gamma_{2-8}$&9.679&81.108&0.01732007&0.11070298&2&8\\
$\Gamma_{1-16}$&9.714&5.4189&0.01139088&0.00108055&1&16\\
\end{tabular}
\end{ruledtabular}
\footnotesize{* Transition with small  transition matrix element and disregarded in this work for CP model analysis comparison.}
\end{table}

\begin{table}
\setlength{\tabcolsep}{1pt}
\caption{\label{tab:baxistransitions} Same as Tab.~\ref{tab:transitions} for polarization parallel axis \textbf{b}.}
\begin{ruledtabular}
\begin{tabular}{ccccc}
Label&$E$ (eV)&$|\mathcal{M}_{cv}|^2$&$c$&$v$\\
\hline
$\Gamma_{1-4}$&5.350&0.06036769&1&4\\
$\Gamma_{1-6}$&5.636&0.14341762&1&6\\
$\Gamma_{1-13}$&7.472&0.00012693&1&13\\
$\Gamma_{2-5}$&8.680&0.02112952&2&5\\
$\Gamma_{4-4}$&9.626&0.00457146&4&4\\
$\Gamma_{3-5}$&9.658&0.00327158&3&5\\
$\Gamma_{4-6}$&9.912&0.00129364&4&6\\
$\Gamma_{2-9}$&9.991&0.08355157&2&9\\
\end{tabular}
\end{ruledtabular}
\end{table}

\subsubsection{Conduction and valence band effective mass parameters}
\begin{figure}
\includegraphics[width=.95\linewidth]{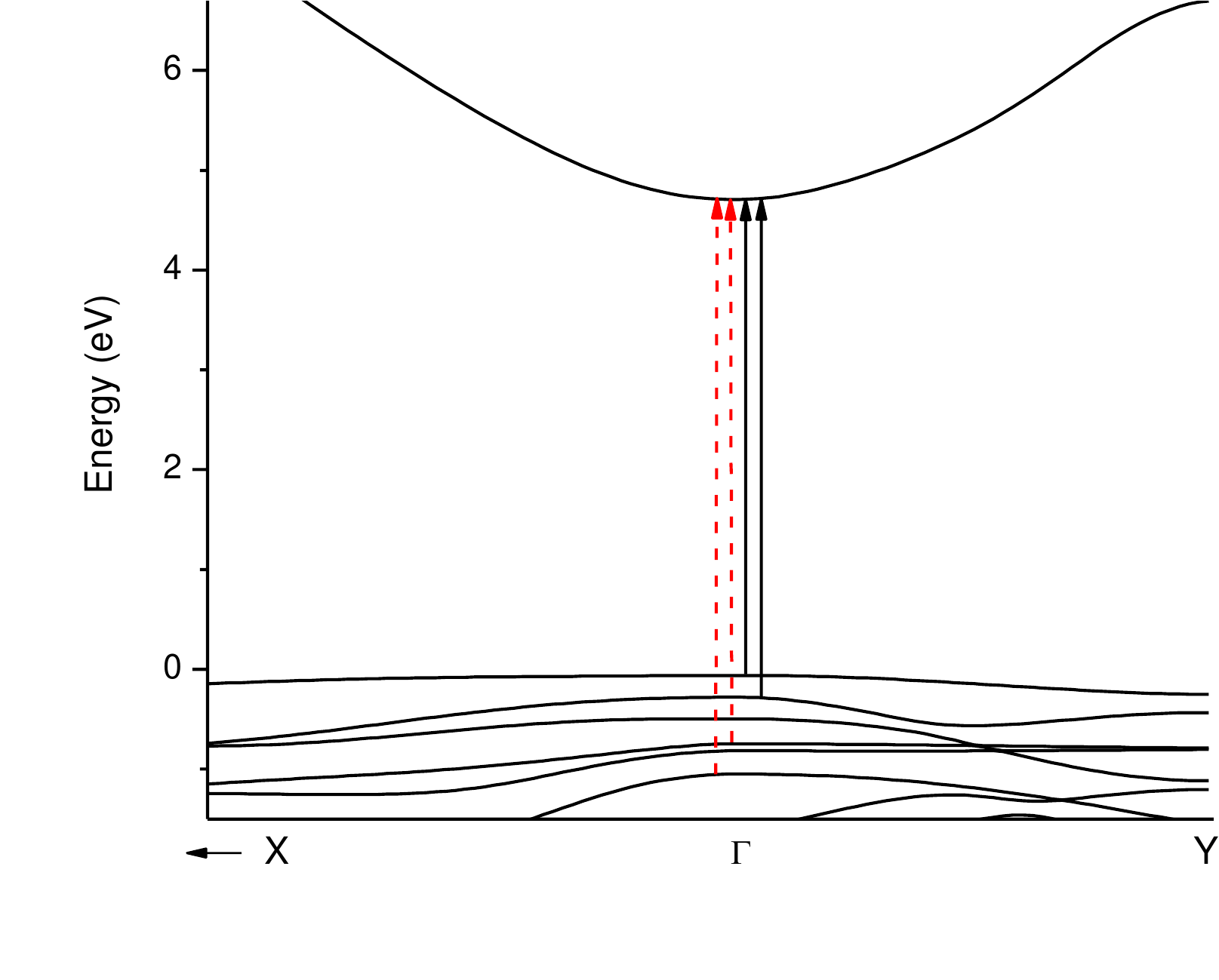}
\caption{\label{fig:gamma} (Color online) Band structure in the vicinity of the $\Gamma$ point. Arrows indicate two lowest vertical band-to-band transitions polarized along the \textbf{b} axis (red dashed arrows) and within the \textbf{a-c} plane (black solid arrows). Note that the reciprocal space direction $\Gamma$-X corresponds to the real space directions parallel to the crystallographic vector \textbf{b}, and the reciprocal space direction $\Gamma$-Y corresponds to the real space direction lying within the \textbf{a-c} plane, inclined at a small angle from the crystallographic vector \textbf{a}.}
\end{figure}

Fig.~\ref{fig:gamma} shows the vicinity of the $\Gamma$ point of the Brillouin zone, with top valence bands and the first conduction band. The four lowest transitions are schematically shown as vertical arrows. The first conduction band is clearly parabolic, and rather symmetric, indicating a nearly isotropic electron effective mass, which is consistent with many previous studies\cite{yamaguchi2004,Furthmuller_2016}. It has been assumed previously\cite{Sturm_2016} that due to the valence bands being generally flat, the hole effective masses are expected to be large, and that the electron effective mass parameter hence dominates the carrier reduced masses for the zone center band-to-band transitions. As can be inferred from Fig.~\ref{fig:gamma}, however, the valence band structure is far from being isotropic. For example, the curvature of the second valence band in the direction $\Gamma-Y$, and the curvature of the fourth valence band in the direction $X-\Gamma$ are similar to the curvature  of the first conduction band, and reveal strong anisotropy as well.

\begin{table*}
\setlength{\tabcolsep}{3pt}
\begin{ruledtabular}
\begin{tabular}{ccccccccc}
$\hat{\mathbf{j}}$ & $m^{\star}_{c1,jj}$ ($m_{\mathrm{e}}$)& $m^{\star}_{v1,jj}$ ($m_{\mathrm{e}}$) & $m^{\star}_{v2,jj}$ ($m_{\mathrm{e}}$) & $m^{\star}_{v4,jj}$ ($m_{\mathrm{e}}$) & $\mu_{calc,jj}$ ($m_{\mathrm{e}}$) & $\varepsilon_{\infty,jj}$ & $R^{\star}_{calc}$ (eV)& $R^{\star}_{exp}$ (eV)\\
\hline
\textbf{a} & 0.224 & 1.769 & 0.466 & 6.649 & 0.151 & 3.474 & 0.171 & 0.23(1)\\
\textbf{b} & 0.301 & $>$10$^a$ & 2.37 & 0.566 & 0.197 & 3.607 & 0.205 & 0.18(1)\\
\textbf{c} & 0.291 & 0.409 & 5.617 & $>$10$^a$ & 0.170 & 3.562 & 0.183 & 0.12(1)\\
\end{tabular}
\end{ruledtabular}
\begin{flushleft}
\footnotesize$^a$Band very flat in this direction
\end{flushleft}
\caption{Effective mass parameters for conduction ($c$) and valence ($v$) bands as indexed, and reduced effective mass parameter for lowest transition along directions $\hat{\mathbf{j}}$ in units of $m_{\mathrm{e}}$. The exciton binding energy $R^{\star}_{calc}$ is then obtained for the lowest transition according to Eq.~\ref{eq:excitonbinding}. $R^{\star}_{exp}$ is obtained from GSE analysis in this work. Data for $\varepsilon_{\infty,jj}$ were obtained from GSE data measured in this work at 0.75~eV.}
\label{effective masses}
\end{table*}

\paragraph{Electron effective mass:} The electron effective mass for $\beta$-Ga$_2$O$_3$ has been studied previously, both by computation and experiment. Computational results consistently predict a very small anisotropy, but span a relatively wide range of values: from (0.12...0.13)$m_e$ (GGA-DFT)\cite{he2006}, through (0.23...0.24)$m_e$ (local density approximation DFT)\cite{yamaguchi2004}, to 0.39$m_{\mathrm{e}}$\cite{Ming-GangJMCA2014}. At the hybrid HF-DFT level the reported values are more consistent: (0.26...0.27)$m_{\mathrm{e}}$\cite{Furthmuller_2016}, (0.27...0.28)$m_{\mathrm{e}}$\cite{varley2010,Peelaerspssb2015Ga2O3meff} with the HSE functional, and 0.34$m_{\mathrm{e}}$ for the B3LYP functional\cite{he2006_2}. Our results (Gau-PBE) are presented in Tab.~\ref{effective masses}, which fall within  this broad range of reported values, but exhibit a slightly higher anisotropy than found in previous studies.

\paragraph{Hole effective mass parameters:} We have analyzed the effective mass parameters for the three valence bands involved in the lowest band-to-band transitions, and data are presented in Tab.~\ref{effective masses}. It is obvious that the hole effective mass anisotropy cannot be neglected for these bands. To our best knowledge, hole effective mass parameters at the $\Gamma$ point have not been reported for $\beta$-Ga$_2$O$_3$ thus far. Yamaguchi\cite{yamaguchi2004} presented values of the top valence band effective mass parameter at point labeled ``E'' away from the zone center and thus not relevant for zone center transitions.

We note an interesting observation from our analysis here: The lowest values of the hole effective mass for each valence band occurs in the approximate  polarization direction of the transition that connects this particular valence band and the lowest conduction band, and which we observe and identify from our GSE and DFT analyses. For the first and topmost valence band, the lowest value of the effective mass occurs along axis \textbf{c}, and the transition $\Gamma_{1-1}$ is polarized nearly along axis \textbf{c} as well. For the second valence band, the lowest effective mass is along axis \textbf{a} and the transition $\Gamma_{1-2}$ is polarized near axis \textbf{a}. For the fourth valence, band the lowest hole effective mass is along \textbf{b} and the transition $\Gamma_{1-4}$ is polarized along \textbf{b}. We thus observe here a clear correlation between the transition selection rules for electronic band-to-band transitions and the values of the carrier effective masses for these transitions. In contrast to previous studies, we find that not only do the hole effective mass parameters matter, but due to their very large anisotropy these parameters may play a decisive role for the polarization of the band-to-band transitions and the associated exciton CP contribution transition parameters.

\subsubsection{Exciton binding energy parameters}
Table~\ref{effective masses} also lists the exciton binding energies for the three lowest band-to-band transitions, for which the excitonic CP model parameters can be  resolved in our experimental data. Exciton binding energy parameters were calculated according to Eq.~\ref{eq:excitonbinding} from reduced mass parameters of each electron and hole pair for the lowest three transitions. Because of the large values of $R^{\star}$ observed in our experiment, which exceed the energy values of all infrared-active phonon modes\cite{Schubert_2016}, we chose the static dielectric function parameters from experimental GSE wavelength-by-wavelength determined epsilon tensor element values at photon energy well above the reststrahlen range, at 0.75~eV (Tab.~\ref{effective masses}). We observe a good agreement between calculated and experimental values for $R^{\star}$. Values for $R^{\star}$ are found different for each transition in contrast to those calculated by Sturm \textit{et al.} (Ref. \onlinecite{Sturm_2016}) who assumed a fixed value of 291~meV for each transition, estimated from an averaged $\varepsilon_{\infty}$ of 3.55 and an electron effective mass parameter of 0.27$m_0$ ignoring hole mass parameter anisotropy.

\subsection{Comparison of DFT, GSE and literature results}

\begin{table*}[!hbt]
\setlength{\tabcolsep}{3pt}
\begin{tabular}{l | c | c | c | c | c | c | c | c | c }
\hline \hline
&\multicolumn{1}{c|}{E$_{ac,0x}$ (eV)} &\multicolumn{1}{c|}{E$_{ac,0}$ (eV)} &\multicolumn{1}{c|}{$\alpha$} &\multicolumn{1}{c|}{E$_{ac,1x}$ (eV)} &\multicolumn{1}{c|}{E$_{ac,1}$ (eV)} &\multicolumn{1}{c|}{$\alpha$} &\multicolumn{1}{c|}{E$_{b,0x}$ (eV)} &\multicolumn{1}{c|}{E$_{b,0}$ (eV)} &\multicolumn{1}{c}{$\alpha$} \\
\hline
This work & 4.92(1) & 5.04(1) & 115.1(1)$^\circ$ & 5.17(1) & 5.40(1) & 25.2(1)$^\circ$ & 5.46(3) & 5.64(1) & \textbf{b}-axis\\
Ref. \onlinecite{Sturm_2015}\footnote{Ellipsometry.} & 4.88 & 5.15\footnote{\label{tabnote:Sturm}A fixed exciton binding energy parameter of 0.27 eV common to all 3 CP transitions listed here was assumed.} & 110$^\circ$ & 5.1 & 5.37$^{\ref{tabnote:Sturm}}$ & 17$^\circ$ & 5.41-5.75 & 5.68-6.02$^{\ref{tabnote:Sturm}}$ & \textbf{b}-axis\\
Ref. \onlinecite{Ricci_2016}\footnote{\label{edge}Room temperature absorption edge.} & - & 4.4 & \textbf{c}-axis & - & 4.57 & \textbf{a}-axis & - & 4.72 & \textbf{b}-axis\\
Ref. \onlinecite{Matsumoto_1974}\footref{edge} & - & 4.54 & \textbf{c}-axis & - & 4.56 & $\perp$ to \textbf{c} and \textbf{b} & - & 4.90 & \textbf{b}-axis \\
\hline
This work DFT & - & 4.740 & 100.504$^\circ$ & - & 4.969 & 7.498$^\circ$ & - & 5.350 & \textbf{b}-axis \\
Ref. \onlinecite{Furthmuller_2016}\footnote{\label{calc}Theory.} & 4.65 & 5.04 & Mainly \textbf{c} & 4.90 & 5.29 & Mainly \textbf{a} & 5.50 & 5.62 & \textbf{b}-axis \\
\hline \hline
\end{tabular}
\caption{Energies and polarization eigenvector directions of the three lowest near-band gap CP transitions including excitonic contributions determined for monoclinic $\beta$-Ga$_2$O$_3$ in this work, in comparison with literature data. The polarization angle $\alpha$ in the \textbf{a-c} plane is defined between axis \textbf{a} and the respective transition dipole polarization direction.}
\label{bandgap}
\end{table*}

\begin{figure*}[hbt]
    \includegraphics[width=.99\linewidth]{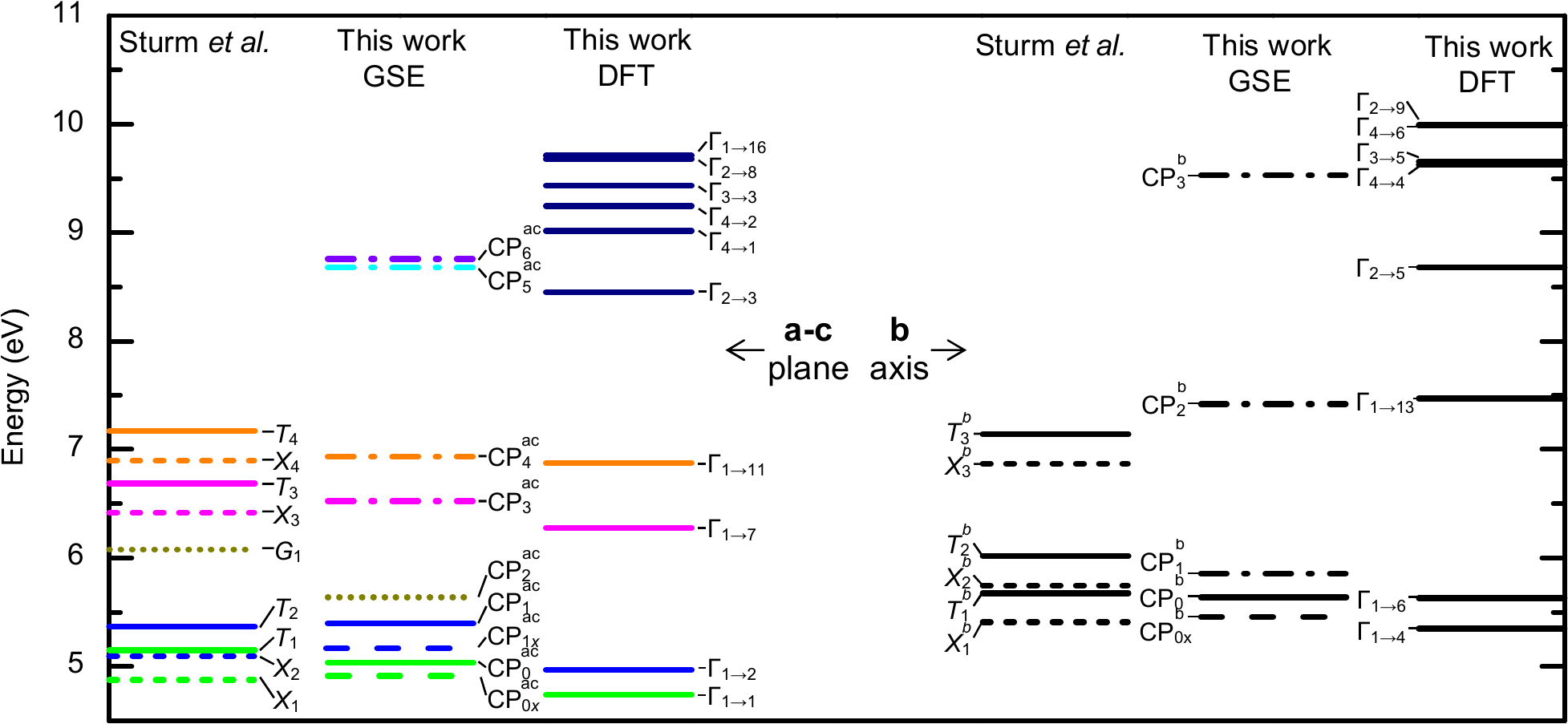}
    \caption{Transition energies determined by CP-MDF analysis and calculated by DFT in our work, in comparison with data reported by Sturm~\textit{et al.} (Ref.~\onlinecite{Sturm_2016}). (Short-dashed (Sturm~\textit{et al.}) and dashed lines (this work):  exciton transitions; solid lines: near-band gap band-to-band transitions; dash-dotted lines:  above-band gap transitions; dotted lines: higher energy transitions. For respective CP-MDF contributions see Sect.~\ref{Eigensection}. DFT levels all refer to band-to-band transitions (solid lines). Color code for DFT \textbf{a-c} plane data are intended to match with order of energy levels identified in GSE CP-MDF analysis.)}
    \label{energies}
\end{figure*}

Figure~\ref{energies} summarizes energy levels below 10~eV determined by CP-MDF analysis and calculated by DFT in our work. Data from Sturm~\textit{et al.} are included for comparison. Overall, the agreement between our GSE and DFT results is excellent, in particular in the near-band-gap transition region, where number of observed transitions (4 in \textbf{a-c} plane, 3 along \textbf{b}) and their energy  levels agree very well. At higher energies, individual transitions identified from DFT cannot be differentiated by GSE analysis, and appear as combined CP contributions.

\paragraph{\textbf{a-c} plane:} CP-MDF and DFT transition energies are listed in Tables.~\ref{parm} and~\ref{tab:transitions}, respectively. The band-to-band transition energy of CP$^{\mathrm{ac}}_{0}$ is found at 5.04~eV, and we identify it with transition $\Gamma_{1-1}$ ($E$=4.74~eV). The second transition, CP$^{\mathrm{ac}}_{1}$, at 5.40~eV, we identify with $\Gamma_{1-2}$ ($E$=4.969~eV). Two further transitions, CP$^{\mathrm{ac}}_{3}$ and CP$^{\mathrm{ac}}_{4}$, are found at 6.53~eV and 6.94~eV, and we identify these with transitions $\Gamma_{1-7}$ ($E$=6.27~eV) and $\Gamma_{1-11}$ ($E$=6.879~eV), respectively. Two above-band gap contributions, CP$^{\mathrm{ac}}_5$ (8.68~eV) and CP$^{\mathrm{ac}}_6$ (8.76~eV), and higher-energy contributions, CP$^{\mathrm{ac}}_{7}$ (10.91~eV) and CP$^{\mathrm{ac}}_{8}$ (12.54~eV) subsume individual contributions from transitions $\Gamma_{2-3}$ ($E$=8.543~eV), $\Gamma_{4-1}$ ($E$=9.0163~eV), $\Gamma_{4-2}$ ($E$=9.2456~eV), $\Gamma_{3-3}$ ($E$=9.432~eV), $\Gamma_{2-8}$ ($E$=9.679~eV), and $\Gamma_{1-16}$ ($E$=9.714~eV). The CP contributions may contain further transitions from different Brillouin zone regions not further investigated. In Fig.~\ref{energies} we also indicate a small contribution (CP$^{\mathrm{ac}}_{2}$ at 6.53~eV) for which we do not observe an equivalent transition in our DFT results. This energy is close to a strong contribution identified along axis \textbf{b}, and it appearance in the \textbf{a-c} plane may originate from lattice defects or from slight experimental misalignment.

\begin{figure}[hbt]
\includegraphics[width=.95\linewidth]{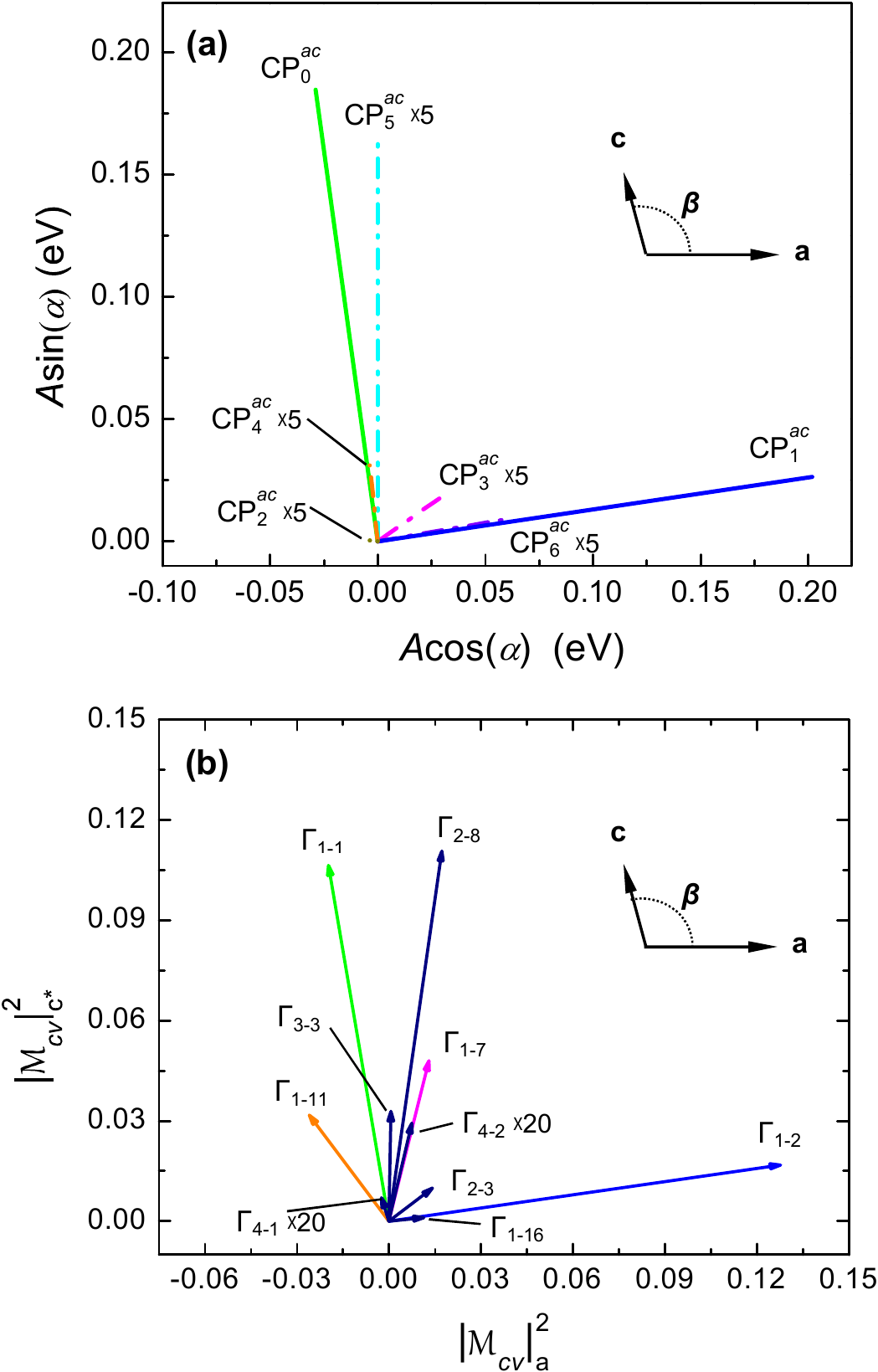}
\caption{Colors and lines styles as in Fig.~\ref{energies}.
(a) Labels as given in Tab.~\ref{parm}. MDF-CP transition eigenvectors, $\hat{\textbf{j}}=\cos\alpha_j\hat{\textbf{x}}+\sin\alpha_j\hat{\textbf{y}}$, multiplied with their respective CP transition amplitude parameter, $A$. The amplitudes of the transitions have been normalized to the amplitude of the first transition, and the CP-MDF unit eigenvectors have been rotated by $\approx$17$^{\circ}$. Small transition amplitude parameters are multiplied for convenience, as indicated. (b) Labels as given in Tab.~\ref{tab:transitions}. DFT calculated transition matrix vector elements, $|\mathcal{M}_{cv}|^2_a$, $|\mathcal{M}_{cv}|^2_{c^{\star}}$. The magnitudes of the transition elements have been normalized by the vector magnitude of the first transition.}
\label{vectors}
\end{figure}

Figure~\ref{vectors} (a), (b) depict CP transition eigenvectors multiplied with their respective CP transition amplitude parameters, or transition matrix element, obtained from GSE and DFT, respectively. Colors and linestyles are as in Fig.~\ref{energies}, for convenient guiding of the eye. As one can see, the agreement between eigen dielectric displacement unit vectors in our CP-MDF approach and the polarization selection conditions obtained from DFT is remarkably good, in particular for the first two band-to-band transitions. None of the identified contributions are purely polarized along either axis \textbf{a}, \textbf{c}, or \textbf{c*}. At higher energies we see a considerable shift between GSE and DFT. We attribute this to an increase in error associated with both the experimental results and the calculations at higher energies. The higher energy transitions predicted by DFT calculations which cannot be resolved from our GSE investigation are shown all in dark blue, the remaining colors correspond to the associated transitions identified by our GSE analysis. Previous work assumed transitions were independently polarized along crystallographic axes. We find here that the lowest two transitions are indeed polarized close to crystal axes \textbf{c} and \textbf{a}, respectively. Matsumoto \textit{et al.} (Ref.~\onlinecite{Matsumoto_1974}) describe the onset of absorption at 4.54~eV and 4.56~eV for polarization along \textbf{c} and for polarization perpendicular to both \textbf{c} and \textbf{b}, respectively, also significantly lower than those found in this work.  Ricci~\textit{et al.} (Ref.~\onlinecite{Ricci_2016}) reported absorption measurements and found the lowest onset occurring with polarization in the \textbf{a-c} plane at 4.5-4.6~eV, which is again at much lower energy than observed in this work. In these previous reports, the exciton contributions were not considered. The closest comparison can be made with the CP-MDF analysis performed by Sturm~\textit{et al.} (Ref.~\onlinecite{Sturm_2016}). We find that the energy levels of the lowest exciton contributions agree very well with those found in our work. However, because Sturm~\textit{et al.} imposed the constraint of fixed and isotropic exciton binding energy parameters, we find in detail different band-to-band transition energy parameters is our work. Table~\ref{bandgap} summarizes energy and polarization eigenvector directions of the near-band-gap transitions determined in this work in comparison with previous reports.

\paragraph{\textbf{b} axis} CP-MDF and DFT transition energies are listed in Tables.~\ref{parmbaxis} and~\ref{tab:baxistransitions}, respectively. Figure~\ref{energies} summarizes transition energies obtained in this work, in comparison with data reported by Sturm~\textit{et al.}. The lowest transition polarized along axis \textbf{b}, CP$^{\mathrm{b}}_{0}$, is found at 5.64~eV, and we identify it with transition $\Gamma_{1-4}$ ($E$=5.350~eV). The second lowest observed transition, CP$^{\mathrm{b}}_{1}$, at 5.86~eV, we identify with $\Gamma_{1-6}$ ($E$=5.636~eV). One further transition, CP$^{\mathrm{b}}_{2}$, is found at 7.42~eV, and we identify this transitions with $\Gamma_{1-13}$ ($E$=7.472~eV). One above-band gap contribution, CP$^{\mathrm{b}}_{3}$ (9.53~eV) and one higher-energy contribution, CP$^{\mathrm{b}}_{4}$ (13.82~eV) account  for contributions from transitions $\Gamma_{2-5}$ ($E$=8.680~eV), $\Gamma_{4-4}$ ($E$=9.626~eV), $\Gamma_{3-5}$ ($E=$9.658~eV), $\Gamma_{4-6}$ ($E$=9.912~eV), and $\Gamma_{2-9}$ ($E$=9.991~eV). The CP contributions may contain further transitions from different Brillouin zone regions not further investigated. Overall, the agreement between GSE and DFT results are very good, in particular for the 3 lowest band-to-band transitions. Matsumoto \textit{et al.} (Ref. \onlinecite{Matsumoto_1974}) using reflectance measurements describe the onset of absorption around 4.9~eV with an absorption edge at 5.06~eV, slightly below our GSE value. Ricci \textit{et al.} (Ref. \onlinecite{Ricci_2016}) reported the onset of absorption occurring at approximately 4.8~eV. Energies reported by Sturm \textit{et al.} (Ref. \onlinecite{Sturm_2016}) are  shifted in detail, which could be explained by the set binding energy constraint.

\subsubsection{Exciton binding energy parameters}

Table~\ref{effective masses} lists exciton binding energy parameters for the 2 lowest transitions within the \textbf{a-c} plane and the lowest transition for axis \textbf{b}. GSE data did not provide sufficient sensitivity to differentiate exciton contributions from higher energy transitions. Using the effective mass parameters obtained from our DFT calculations and DFT calculated transition energies, we derived binding energies for Wannier-type excitons, and assuming absence of phonon interaction. We observe that both data sets reflect direction dependent binding energy parameters, and both sets show similar magnitude. Our calculated data predict the largest [smallest] binding energy parallel axis \textbf{b} (0.205~eV) [\textbf{a} (0.171~eV)], while our GSE results results indicate axis \textbf{a} (0.23~eV) [\textbf{c} (0.12~eV)]. The deviations may originate from the simplification made by assuming an isotropic exciton picture separately for each transition. Nonetheless, our results clearly show that the exciton contributions should not be assumed direction independent, and further theoretical and experimental analysis for semiconductor materials with monoclinic crystal symmetry may be needed. Our finding further suggests that the assumed binding energies of 270~meV by Sturm \textit{et al.} overestimated the exciton shift of the onset of absorption in $\beta$-Ga$_2$O$_3$.

\section{Conclusions}

The eigen dielectric displacement model, previously described for analysis of long wavelength generalized ellipsometry data as well as for  near-infrared to ultra violet spectral regions for for $\beta$-Ga$_2$O$_3$, was applied for an analysis expanding into the vacuum ultra violet spectral region. In our analysis we permitted for direction dependent exciton binding parameters. We differentiated 9 critical point contributions in the \textbf{a-c} plane, the lowest 2 of which were modeled with excitonic contributions. Additionally, we observed five critical point contributions in the \textbf{b} direction, with an excitonic contribution associated with the lowest transition. The binding energy parameter of these excitons associated with the lowest band-to-band transitions along each of the experimental coordinate system axes were shown to present with distinct values ranging from 0.12-0.23~eV. Additionally, transitions in the monoclinic plane, which does not contain any symmetry operation, were found to be distributed within the plane and none of them would align with major crystal directions \textbf{a} or \textbf{c}. Our experimental analysis compares well with results from density functional theory calculations performed using a recently proposed Gaussian-attenuation-Perdue-Burke-Ernzerhof density functional, We presented and discussed the order of the fundamental direct band-to-band transitions and their polarization selection rules, the electron and hole effective mass parameters for the three lowest band-to-band transitions, and their exciton binding energy parameters, in excellent agreement with our experimental results. We find that the effective masses for holes are highly anisotropic and correlate with the selection rules for the fundamental band-to-band transitions, where the observed transitions are polarized closely in the direction of the lowest hole effective mass for the valence band participating in the transition. The MDF approach and parameter set for $\beta$-Ga$_2$O$_3$ presented here will become useful for ellipsometry analysis of heterostructures, and may be expanded for description of alloys with monoclinic crystal symmetry.

\section{Acknowledgments}

We dedicate this work to our friend, colleague, and educator Professor Erik Janz\'en, who left us untimely and unexpected. We would like to thank J\"urgen Furthm\"uller for useful discussions. This work was supported by the National Science Foundation (NSF) through the Center for Nanohybrid Functional Materials (EPS-1004094), the Nebraska Materials Research Science and Engineering Center (DMR-1420645), the Swedish Research Council (VR2013-5580), and the Swedish Foundation for Strategic Research (SSF, FFL12-0181 and RIF14-055). Partial financial support from NSF (CMMI 1337856, EAR 1521428), and J.~A.~Woollam Foundation is also acknowledged. DFT calculations were performed using the computing resources at the Center for Nanohybrid Functional Materials and at the Holland Computing Center at the University of Nebraska-Lincoln.

\bibliography{Ga2O3_ref}
\end{document}